\documentclass[traditabstract]{aa}  
\usepackage[varg]{txfonts}
\usepackage{graphicx}
\usepackage{subfigure}
\usepackage{rotating}
\usepackage{color}
\usepackage{ragged2e}
\usepackage{soul} 
\usepackage{float}
\usepackage{amsmath} 
\usepackage{wrapfig,lipsum}

\usepackage[flushleft]{threeparttable} 

\usepackage{soul}

\begin{document}
\defcitealias{2020A&A...641A.171M}{M20}

\title{The origin of double-peak emission-line galaxies: \\rotating discs, bars or galaxy mergers?}
\titlerunning{The mechanisms behind double-peak emission-line galaxies}
\authorrunning{Maschmann et al.}

\author{Daniel Maschmann\inst{1, 2}, Ana\"elle Halle\inst{1}, Anne-Laure Melchior\inst{1}, Francoise Combes\inst{1, 3}, Igor V. Chilingarian\inst{4,5}\\ 
}
\institute{Sorbonne {Universit\'e},
LERMA, Observatoire de Paris, PSL university, CNRS, F-75014, Paris, France\\
\email{Daniel.Maschmann@observatoiredeparis.psl.eu, anaelle.halle@observatoiredeparis.psl.eu}
\and
Steward Observatory, University of Arizona, 933 N Cherry Ave,Tucson, AZ 85721, USA
\and
Coll\`ege de France, 11, Place Marcelin Berthelot, F-75005, Paris, France
\and
Center for Astrophysics -- Harvard and Smithsonian, 60 Garden St. MS09, Cambridge, MA, 02138, USA 
\and
Sternberg Astronomical Institute, M.V. Lomonosov Moscow State University, 13 Universitetsky prospect, Moscow, 119991, Russia
}
\date{Received 26 July 2022/ accepted }
\abstract {
Emission lines with a double-peak (DP) shape, detected in the centre of galaxies, have been extensively used in the past to identify peculiar kinematics such as dual active galactic nuclei (AGN), outflows or mergers. With a more general approach considering a large DP galaxy sample selected from the SDSS, a connection to minor merger galaxies with ongoing star formation was suggested.
To gain a better understanding of different mechanisms creating a DP signature, we here explore synthetic SDSS spectroscopic observations computed from disc models and simulations.
We show how a DP signature is connected to the central part of the rotation curve of galaxies, which is mostly shaped by the stellar bulge. 
We, furthermore, find that bars can create strong DP emission-line signatures when viewed along their major axis. 
Major mergers can form a central rotating disc in late post-coalescence merger stages (1\,Gyr after the final coalescence), which creates a DP signature. Minor mergers tend to show a DP feature with no correlation to the galaxy inclination within 350\,Myr after the final coalescence.
Comparisons of these scenarii with observations disfavour major mergers, since these show predominantly elliptical and only a few S0 morphologies. Furthermore, at such a late merger stage the enhanced star formation is most likely faded.
Bars and minor mergers, on the other hand, can be compared quite well with the observations. Both observations coincide with increased star formation found in observations, and minor mergers in particular do not show any dependency with the observation direction.
However, observations resolving the galaxy kinematics spatially are needed to distinguish between the discussed possibilities. More insight into the origin of DP will be gained by a broader comparison with cosmological simulations. The understanding of the DP origin can provide important tools to study the mass growth of galaxies in future high redshift surveys.
}

\maketitle
\section{Introduction}\label{sect:introduction}
The evolution of galaxies involves dynamical processes such as galaxy mergers whose frequency remain difficult to measure over cosmic time. Studies based on photometry may for example not always be efficient at identifying these processes, while kinematics may be misleading. Mergers have been extensively studied using simulations \citep[e.g.][]{1972ApJ...178..623T,1985ARA&A..23..147A, 1995ApJ...448...41H,2005A&A...437...69B,2007A&A...468...61D,2010MNRAS.404..575L} and observations \citep[e.g.][]{1994A&A...281..725C, 2003A&A...405...31B,2004AJ....128..163L,2005AJ....130.1516D, 2008AJ....135.1877E, 2013MNRAS.435.3627E}, resulting in a good understanding on how galaxy mergers can fuel star formation, trigger active galactic nuclei (AGN) and transform the morphology of galaxies. 

Especially studies dealing with different stages of galaxy merger rely on an accurate identification of mergers. Interacting galaxies can be identified through their projected separation \citep{2005AJ....130.1516D,2008AJ....135.1877E,2011MNRAS.412..591P}. Major mergers in an early phase of their coalescence show strong tidal features and can be identified though their perturbed morphology \citep[e.g.][]{2004AJ....128..163L}. 
After the final coalescence, tidal features and perturbations will gradually fade and it becomes increasingly difficult to correctly distinguish between post-merger galaxies and isolated galaxies. From hydrodynamical simulations, major (resp. minor) mergers can be identified after $\sim 200-400$\,Myr (resp. 60\,Myr) using photometric diagnostics \citep{2010MNRAS.404..575L}. Using a combination of several photometric classifiers to a linear discriminant analysis, \citet[][]{2019ApJ...872...76N} succeeded in identifying galaxy mergers over a merger timescale of 2\,Gyr. Including stellar kinematics measured with integrated field spectroscopic observations, \citet{2021ApJ...912...45N} increased the detection sensitivity for post-coalescence mergers. However, it remains challenging to apply these techniques to observations and identify post-coalescence mergers.

As predicted in \citet[][]{1980Natur.287..307B}, the two super-massive black holes of the progenitors of a merger will eventually merge in the course of the coalescence. Previous to this event, the two nuclei will stay at a separation $>1\,{\rm kpc}$ for $\sim\,100\,{\rm Myr}$. When both nuclei are AGNs, it is possible to observe this phenomenon using telescope providing high enough resolution. Such dual AGNs were observed using X-ray observation \citep{2003ApJ...582L..15K}, radio observations \citep{2004ApJ...602..123M, 2006ApJ...646...49R} and long-slit spectroscopy, revealing a double-peak (DP) signature \citep{2007ApJ...660L..23G}.
The connection between the kinematic footprint and a dual AGN was further discussed in \citet{2009ApJ...698..956C}. Systematic studies on DP emission-line AGNs using additional high resolution observations were able to distinguish between dual AGNs, AGN driven outflows or rotating discs \citep{2011ApJ...737L..19C,2014ApJ...789..112C,2015ApJ...806..219C,2018ApJ...867...66C, 2015ApJ...813..103M, 2016ApJ...832...67N}.

In general a DP emission-line profile traces  multiple line-of-sight velocities. AGNs are compact and bright sources and therefore dual AGNs, moving at two different velocities, are particularly interesting to study late stages of mergers. \citet{2012ApJS..201...31G} built up a DP-galaxy sample, including also non-AGNs and gathered 3\,030 galaxies, of which only 30\,\% are classified as AGNs.
These DP emission-line signatures can have various causes: a compact rotating disc, gas outflow or inflow, two nuclei or the alignment of two galaxies inside the line-of-sight. 
In \citet{2020A&A...641A.171M} (hereafter \citetalias{2020A&A...641A.171M}) 5\,663 DP emission-line galaxies were selected using an automated selection procedure. Interestingly, only 14\,\% were found to be AGNs. Different scenarii were discussed to explain the origin of DP emission lines and a recent minor merger was favoured as the underlying process. As these results are particularly relevant for this work, the main findings are explained in detail in Sect.\,\ref{ssect:summary_m20}. On the one hand, it is still challenging to conclude on the origin of DP emission lines for an individual galaxy, relying only on one optical spectrum and a snapshot. On the other hand,
a merger scenario becomes increasingly likely if one finds different characteristics in the two emission-line components \citep{2019A&A...627L...3M}. Using integrated field spectroscopy, \citet{2021A&A...653A..47M} detected two galaxies aligned inside the line-of-sight, creating a DP emission line. 
In a recent study, the molecular gas content of DP galaxies selected from above the main star-forming sequence was studied in \citet{2021arXiv211212796M}. 20\,\% of the DP galaxies show the same kinematic feature in the CO emission line distribution which traces the molecular gas, indicating highly concentrated gas reservoir. Furthermore, in nearly all galaxies, a central star formation enhancement was found, and 50\,\% of the sample were identified as visual mergers or showed tidal features. Taking into account that the observed galaxies have a significantly larger molecular gas reservoir than expected for galaxies situated above the main sequence, the most plausible explanation of the DP emission line profile was found to be a recent minor merger which funnelled gas into the central regions and fuels a compact star-formation region.   

To better understand the observed DP emission-lines, we here use models and simulations of galaxies. 
We investigate possible origins of DP emission lines in this work and determine under which conditions a DP signature may be detected in isolated galaxies, ongoing mergers and post-mergers. More precisely, we seek to identify DP emission lines in the conditions of observations with a SDSS-like $3^{\prime\prime}$ spectroscopic fibre observations centred on the brightest region of the targeted system. We study the connection between identified DP signatures in the line-of-sight and the kinematic processes inside the observed systems. 

In Sect.\,\ref{sect:rotating_discs}, we describe axisymmetric models of disc galaxies and then study numerical simulations of such galaxies in which non-axisymmetric patterns, especially bars (in the central regions of interest), form. In Sect.\,\ref{sect:merger}, we characterise major and minor-merger simulations and identify under which circumstances a DP emission line can be detected. We then discuss in Sect.\,\ref{sect:discussion} the found results in the context of past work on DP emission line galaxies and conclude in Sect.\,\ref{sect:conclusion}. In this work, a cosmology of $\Omega_{m} = 0.3$, $\Omega_{\Lambda} = 0.7$ and $h = 0.7$ is assumed.

\section{Observations of double-peak emission-line galaxies in the SDSS}\label{sect:dp_sdss}
The focus of this work is to determine the origin of DP emission-line profiles. To accomplish this, we analyse synthetic emission-line spectra from galaxy models and galaxy simulations. To frame this analysis in the context of observations, we  here recapitulate the results of \citetalias{2020A&A...641A.171M} and summarise the most important sample characteristics of their assembled DP galaxy sample. We then select three redshift values in order to represent the redshift distribution of the DP sample found in \citetalias{2020A&A...641A.171M} and describe how to detect DP profiles in synthetic emission-line spectra.

\subsection{Double-peak detection in \citetalias{2020A&A...641A.171M}}\label{ssect:summary_m20}
The selection procedure of \citetalias{2020A&A...641A.171M} is divided into multiple stages which make use of emission-line parameters provided by the Reference Catalogue of Spectral Energy Distribution (RCSED) \citep{2017ApJS..228...14C}. 
In a first step, galaxies with a high enough signal-to-noise ratio of ${\rm S/N} > 10$ in either the H$\alpha$ or the [OIII]$\lambda 5008$ emission lines were selected. Then, galaxies with emission-lines which are better described by a non-parametric fit than by a single-Gaussian fit were selected and all emission-lines with a ${\rm S/N} > 5$ were stacked. 
The resulting emission-line profile was fitted by both a single and a double-Gaussian function. Relying on an F-test of the two fits, an amplitude ratio threshold of the two double-Gaussian components, and a minimal threshold in velocity difference $\Delta v_{\rm DP} > 3 \delta v$, with $\delta v$ the SDSS bin-width of $69\,{\rm km\,s^{-1}}$, 7\,479 DP-candidates were selected. 
In a second stage, each emission line was individually fitted with a single and a double-Gaussian fit. The double-Gaussian fit is restrained to the parameters found from the stacked emission line, however, the parameters can still vary within their uncertainties. All emission lines with a ${\rm S/N} > 5$ were flagged as a DP emission line if they satisfy the following conditions: (1) the reduced chi-square value of the double Gaussian fit must be smaller than the value for the single Gaussian fit, (2) the double-Gaussian amplitude ratio $A_{1}/A_{2}$ must fulfill the condition ${1}/{3} < A_{1}/A_{2} < 3$, and (3) each of the double-Gaussian emission-line component must be detected with at least ${\rm S/N} > 3$. 
In a third stage, galaxies were selected with a DP in their strongest emission lines, resulting in a final sample of 5\,663 DP galaxies.

In order to compare the selected DP sample to galaxies with only a single peaked (SP) emission-line profile,  a no-bias-control-sample was selected with the same emission-line S/N properties, redshift distribution and stellar mass distribution as the DP sample. 
Analysing the morphology of these two samples, the same visual merger rate was found between DP and SP galaxy. However, DP galaxies are more likely to be classified as S0 galaxies (36\,\%) in comparison to SP galaxies (20\,\%). Furthermore, DP galaxies classified as spiral galaxies tend to have larger bulges and are more likely classified as Sa or Sb galaxies whereas SP galaxies tend to be classified as Sc and Sd. 
A detailed analysis of the spectroscopic kinematics revealed a significant higher stellar velocity dispersion in DP galaxies in comparison to SP galaxies. A correlation between the galaxy inclination and the gas kinematics was found for SP galaxies, but not for DP galaxies. DP galaxies also deviate from the Tully-Fisher relation in contrast to SP galaxies. When considering each individual fit component of the DP sample, however, a good agreement with the Tully-Fisher relation is found.
Considering star-forming galaxies, a central star-formation enhancement was found for DP galaxies but not for SP galaxies. 
Conclusively, these observations agree in particular with a model of repetitive minor mergers which effectively transport gas into the central regions and drive bulge growth as described in \citet{2007A&A...476.1179B}.

\subsection{SDSS spectroscopic measurements at different redshifts}\label{ssect:fibre_size}
\begin{figure}
\centering 
\includegraphics[width=0.48\textwidth]{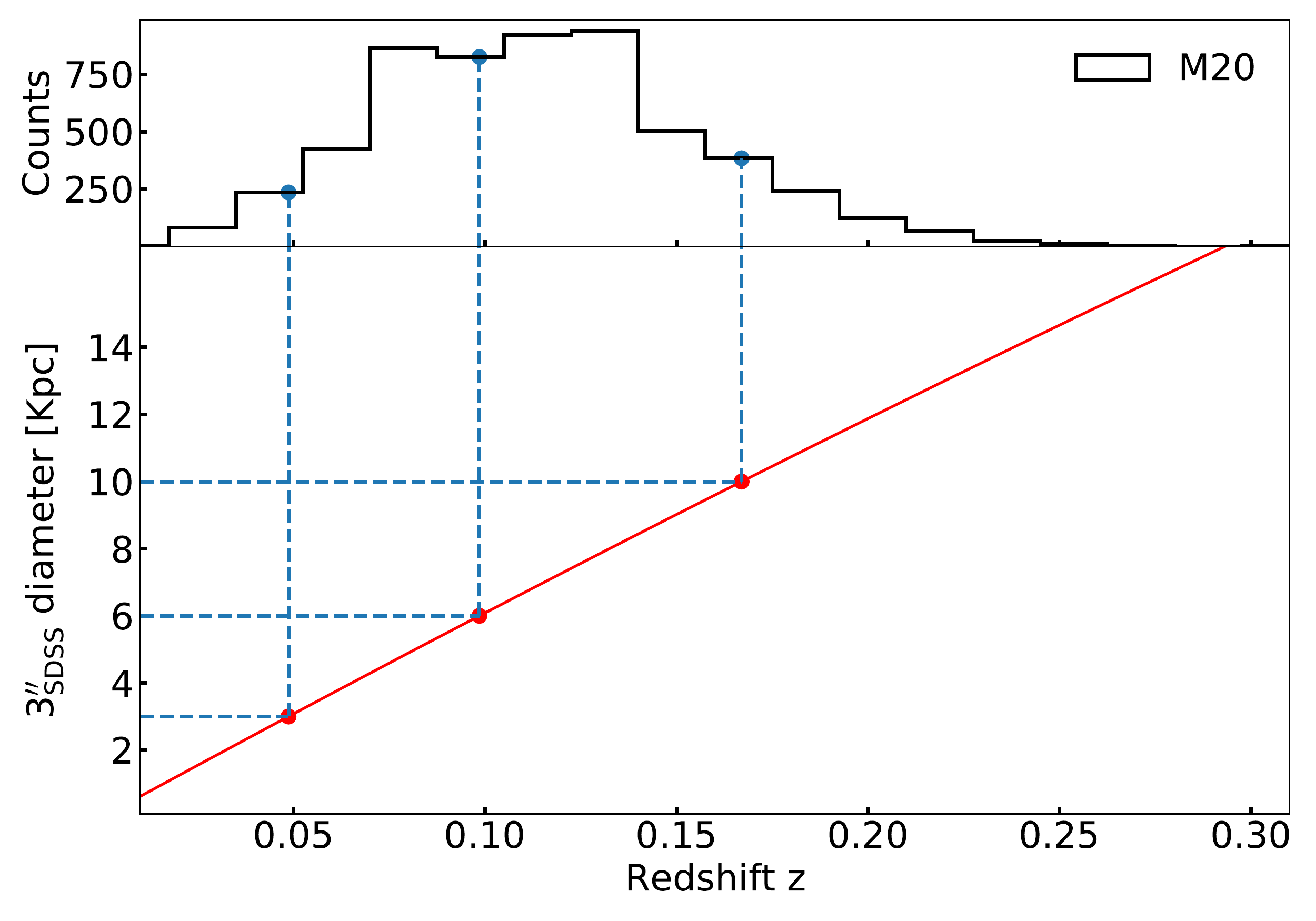}
\caption{Redshift distribution of the DP galaxy sample  \citepalias{2020A&A...641A.171M} (top panel) and the conversion curve between redshift and the fibre diameter in kpc of the SDSS 3$^{\prime\prime}$ (red curve in the bottom panel). We mark with blue dashed line the three representative redshifts and corresponding fibre diameters: $z=0.05$, $z=0.1$ and $z=0.17$ corresponding to a fibre diameter of 3, 6 and 10\,kpc, respectively.}
\label{fig:fibre_size}%
\end{figure}
The spectroscopic observation in the SDSS is taken within a 3$^{\prime\prime}$ region centred on the brightest spot of a galaxy \citep{2009ApJS..182..543A}. Hence, this spectrum probes the central 0.6\,kpc in low redshift galaxies at $z=0.01$ and 30\,kpc for the most distant spectroscopic observations in the SDSS at about $z=0.55$. In the latter case, the SDSS spectrum probes roughly the entire galaxy, whereas for a nearby galaxy the spectroscopic measurement probes only the very centre.
In Fig.\,\ref{fig:fibre_size}, we show the redshift distribution of the DP galaxy sample of \citetalias{2020A&A...641A.171M} and a conversion curve between the fibre diameter and the redshift. The DP sample has a median redshift of $z=0.11$ and 99\,\% of the sample has a redshift of $z < 0.22$. Only 57 galaxies are situated at higher redshift up to a value of $z=0.34$. In order to represent this distribution, we select three representative redshift values: $z=0.05$, $z=0.1$, and $z=0.17$, corresponding to a fibre diameter of 3, 6, and 10\,kpc, respectively. In the following, we will analyse simulated SDSS spectral observations of analytical models and galaxy simulations with these three fibre diameters. 

\subsection{Double-peak detection in synthetic emission-line spectra}\label{ssect:synthetic_dp}
In order to test whether a computed emission-line profile from an axisymmetric model or a galaxy simulation shows a double-peak feature, we develop a detection algorithm similar to \citetalias{2020A&A...641A.171M}. 
In a first step, we convolve the produced line-of-sight velocity profiles with the mean instrumental broadening of $61\,{\rm km\,s^{-1}}$ \citepalias{2020A&A...641A.171M} from the SDSS spectral detector, and compute the resulting signal with the SDSS bin-width of $\delta v=69\,{\rm km\,s^{-1}}$. We then fit a single and a double Gaussian function to the velocity profile and select DP galaxies satisfying the following criteria: 
\begin{enumerate}
	\item $\chi^2_{\nu}({\rm single}) > \chi^2_{\nu}({\rm double})$\
	\item ${1}/{3} < A_{1}/A_{2} < 3$
    \item $\Delta v_{\rm DP} = |\mu_2 - \mu_2| > 3\,\delta v$ 
\end{enumerate}
where $\chi^2_{\nu}({\rm single})$ (resp. $\chi^2_{\nu}({\rm double})$) is the reduced chi-square computed for the single (resp. double) Gaussian fit, $A_{1}$ and $A_{2}$ are the amplitudes of the two Gaussian functions in the double Gaussian fit and $\Delta v_{\rm DP}$ is the velocity difference between the blue and redshifted component. In a first step of selection of DP candidates in \citetalias{2020A&A...641A.171M}, an F-test was used. However, this was mostly motivated to distinguish a DP from a SP profile in the case of a noisy spectra. Since we do not include noise in our synthetic emission-line profiles, we only use the chi-square ratio as such a selection criterion.

\section{Rotating discs}\label{sect:rotating_discs}
Double-peaked emission lines can be due to the rotation of discs. In order to investigate when such a detection of DP is possible, we first construct an idealised galaxy model with an axisymmetric rotating gas disc. We modify the rotation curve of the model by varying the mass concentration of a stellar bulge and study the resulting gas line-of-sight velocity distribution. We also study the effect of a change in the concentration of the gas density profile. Using simulations of isolated galaxies, we then investigate how the presence of a bar may impact the detection of a DP signature. 

\subsection{Axisymmetric models}\label{ssect:axi_sym_model}
\begin{table}
\caption{Mass and length parameters for the Sa galaxy.}
\label{table:gSa_params}
\begin{tabular}{ccccc}
$M_{\rm gas}$ & $M_{\rm * \, disc}$ & $M_{\rm * \, bulge}$ & $M_{\rm DM}$ &  $[2.3\times 10^9 M_{\odot}]$\\
\hline
4 & 40 & 10 & 50 & 
\end{tabular}
\begin{tabular}{ccccccc}
$a_{\rm gas}$ & $h_{\rm gas}$  & $a_{\rm * \, disc}$ & $h_{\rm * \, disc}$ & $b_{\rm * \, bulge}$ & $b_{\rm DM}$ & [kpc] \\
\hline
5 & 0.2 & 4 & 0.5 & 0.2-3 & 10 & 
\end{tabular}
\end{table}
\begin{figure}
\centering 
\includegraphics[width=0.48\textwidth]{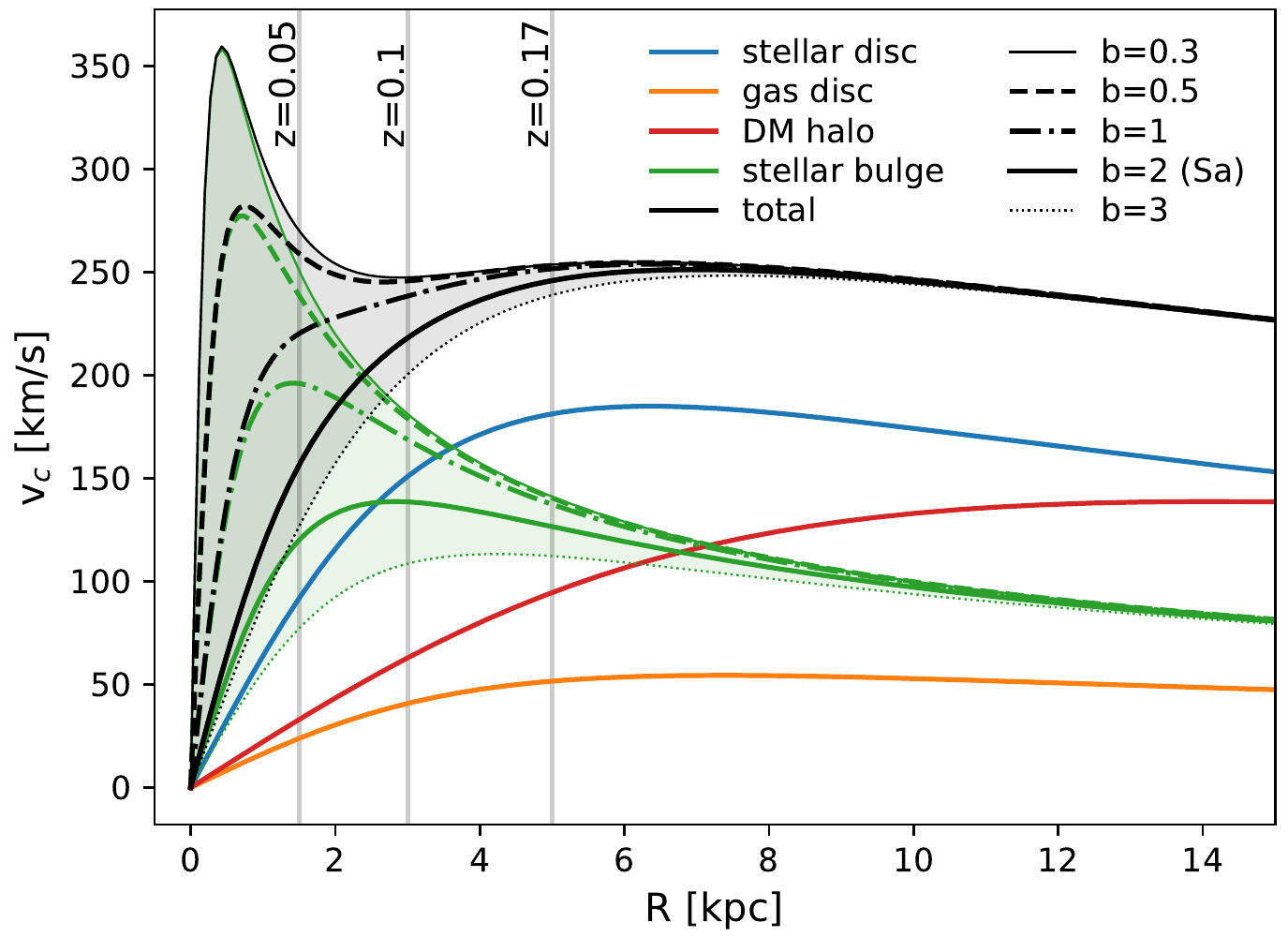}
\caption{Rotation curves of a disc galaxy for different characteristic radii of the stellar bulge. We show with coloured lines the contribution of each component, and in black the total rotation curve. In blue (resp. orange), we show the contributions of the stellar (resp. gaseous) disc, described by Miyamoto-Nagai density profiles. With a green (resp. red) line, we show the contributions of the stellar bulge (resp. dark-matter halo), described by Plummer density profiles. We show with different line styles the contribution of the bulge and the total rotation curve for bulges with different characteristic radii $b$. A characteristic bulge radius $b=2$ (thick solid green and black lines) corresponds to the fiducial Sa galaxy. Vertical grey lines are plotted at the radii of the simulated fibre for the redshifts $z=0.05$, 0.1 and 0.17.}
\label{fig:rot_curves}%
\end{figure}
\begin{figure}
\centering 
\includegraphics[width=0.48\textwidth]{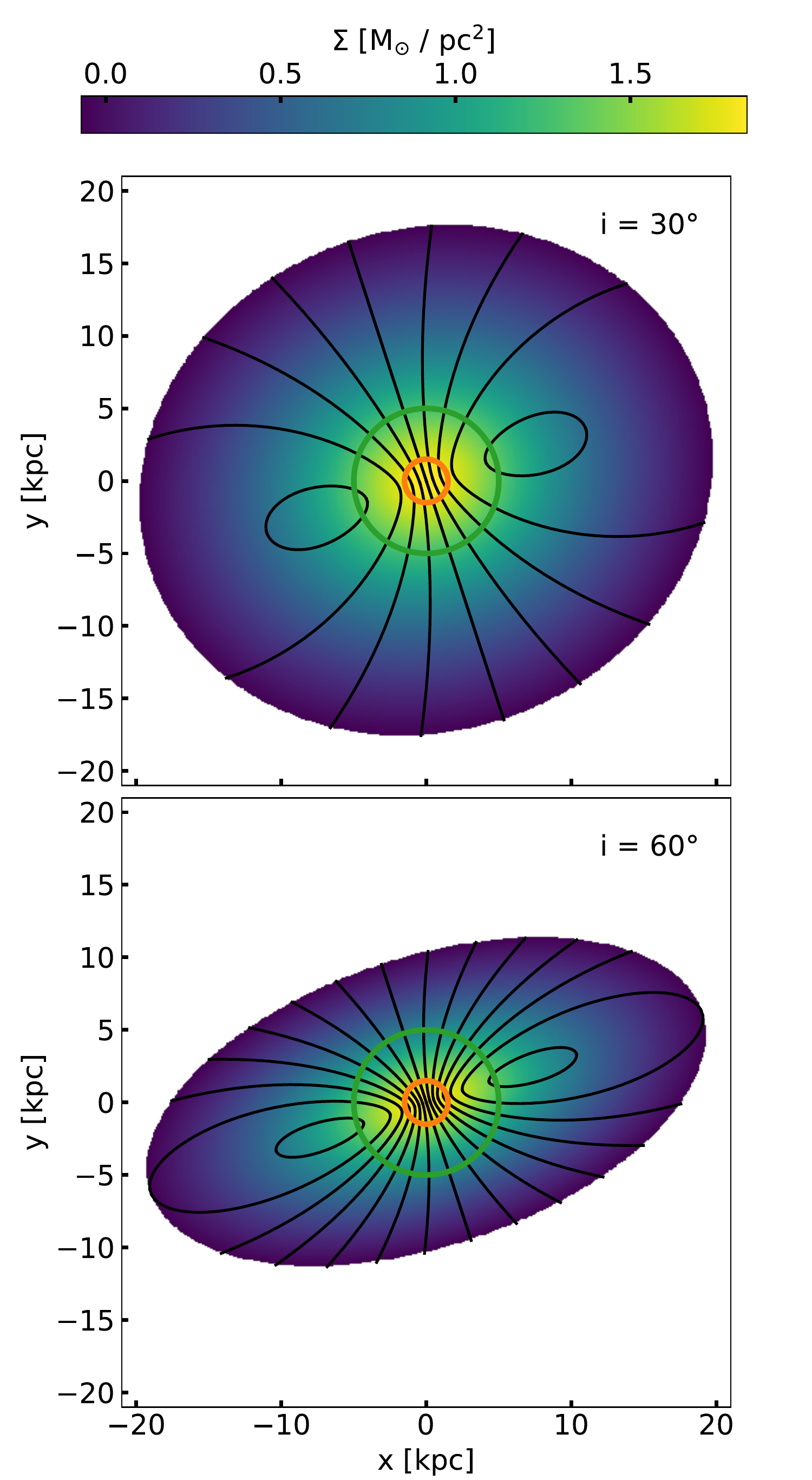}
\caption{Gaseous disc model at two different inclinations. We show 2D projections of a Miyamoto-Nagai density profile as described in Equation\,\ref{eq:mn} at an inclination of $i=30^{\circ}$ (top panel) and $i=60^{\circ}$ (bottom panel). Both discs are turned by a position angle of $20^{\circ}$. The colour-bar indicates the surface density. With black lines, we indicate iso-velocity curves, with velocity values separated by 30\,km\,s$^{-1}$ (with a value of 0\,km\,s$^{-1}$ on the minor axes). With an orange (resp. green) circle, we show the area observed by a $3^{\prime\prime}$ spectroscopic fibre at a redshift of $z=0.05$ (resp. $z=0.17$). The more inclined the disc, the smaller the distance of equidistant velocity lines and thus the steeper the velocity gradient probed by the spectroscopic fibre.}
\label{fig:equi_vel}%
\end{figure}
The model, for its fiducial set of parameters, reproduces an Sa galaxy. Potential-density pairs are used for all four components. The gas and stellar discs each have a Miyamoto-Nagai density profile \citep{1975PASJ...27..533M}:
\begin{equation}\label{eq:mn}
\rho_d(R,z) = \left( \dfrac{h^2 M}{4 \pi} \right) \dfrac{a R^2 + (a + 3\sqrt{z^2 + h^2}) (a + \sqrt{z^2 + h^2})^2}
    {\left[ a^2 + (a + \sqrt{z^2 + h^2})^2\right]^{\frac{5}{2}} (z^2 + h^2)^{\frac{3}{2}}},
\end{equation}
where $M$ is the total mass of the disc, $a$ is a radial scale length, and $h$ is a vertical scale length. The stellar bulge and the dark-matter halo each have a Plummer profile \citep[][pag. 42]{1987gady.book.....B}:
\begin{equation}\label{eq:plummer}
    \rho_s (r) = \left(\frac{3 M}{4 \pi r^3}\right)  \left(1 + \frac{r^2}{b^2}\right)^{-\frac{5}{3}},
\end{equation}
where $M$ is the total mass of the component and $b$ a characteristic radius. The profile parameters for the four components are given in Table~\ref{table:gSa_params}, for an Sa galaxy.

The rotation curve is shown on Fig.~\ref{fig:rot_curves} (thick black curve for an Sa), with the detail of the contributions of the different components. The individual contribution of each disc component is $\sqrt{ R  \dfrac{\partial \Phi_d}{\partial R} \bigg\rvert_{z=0}}$ with $\Phi_d$ the gravitational potential of the disc component: 
\begin{equation}
    \Phi_d(R, z) = - \dfrac{G M}{\sqrt{R^2 + (a+\sqrt{z^2 + h^2})^2}},
\end{equation}
and the individual contribution of each spherical component (stellar bulge or dark-matter halo) is $\sqrt{ r \dfrac{\partial \Phi_s}{\partial r}}$ with $r$ the spherical radius and $\Phi_s$ the gravitational potential of the spherical component: 
\begin{equation}
    \Phi_s(r) = - \dfrac{G M}{\sqrt{r^2 + b^2}}.
\end{equation}
The rotation curve is then obtained as the square root of the quadratic sum of the four contributions. For such an Sa galaxy, the bulge dominates the rotation curve in the central parts, creating a steep rise of the rotation curve at small galactocentric radii (see the thick green curve representing the bulge contribution on Fig.~\ref{fig:rot_curves}). 

\subsubsection{Emission-lines of a fiducial Sa galaxy}\label{ssect:fiducial_sa}
\begin{figure*}
\centering 
\includegraphics[width=0.98\textwidth]{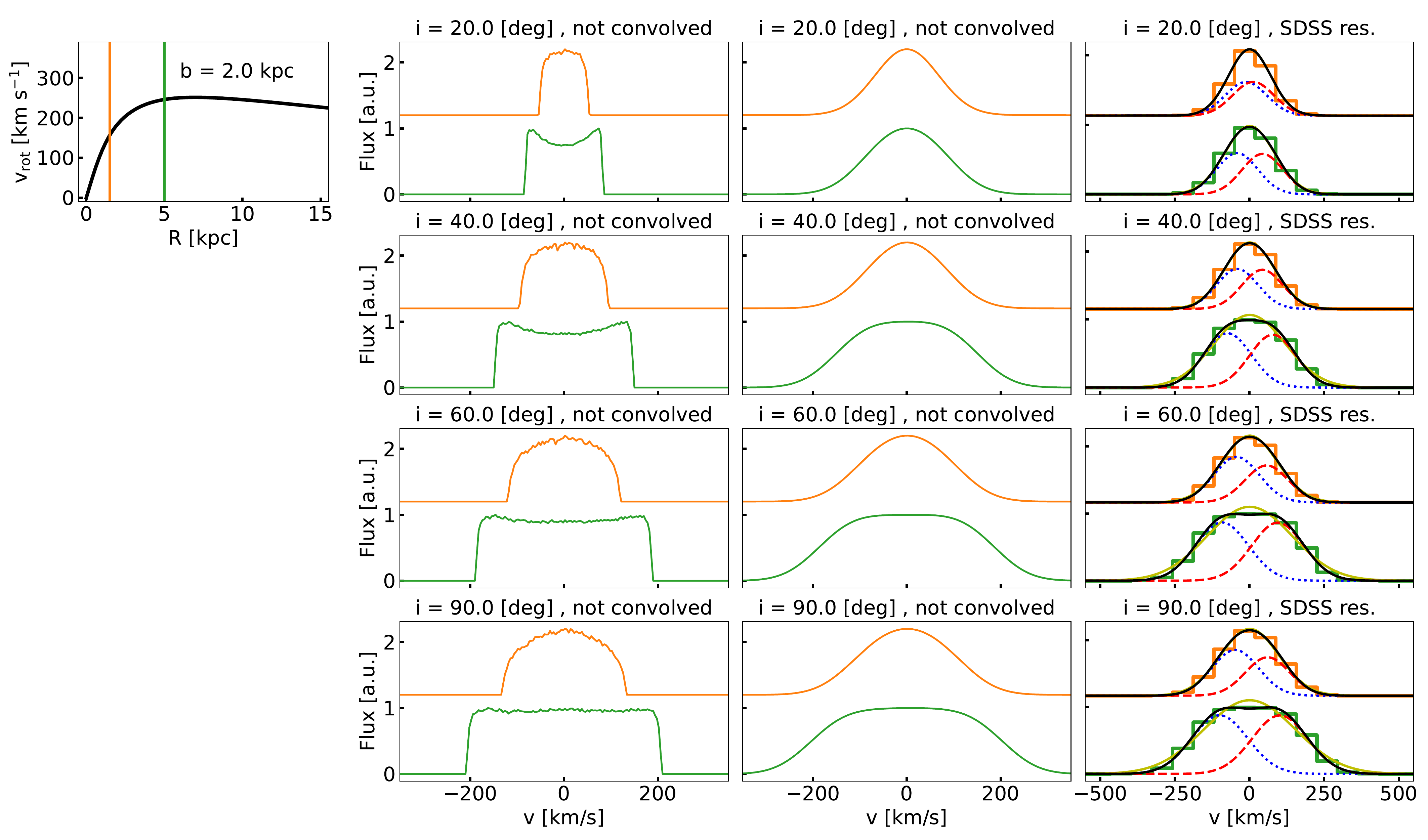}
\caption{Emission-line profiles of gaseous-disc model observed at different inclinations. On the top left, we show a rotation curve calculated by a model including a stellar and gaseous disc, a stellar bulge and a dark-matter halo, parametrised as summarised in Table\,\ref{table:gSa_params} with $b=2.0\,{\rm kpc}$. We compute the emission-line profiles observed within a $3^{\prime\prime}$ spectroscopic fibre for two redshift values: $z=0.05$ and $z=0.17$. The region probed by these observations are marked by orange and green line, respectively. For the emission-line profiles we also show the two spectroscopic observations in green and orange, with an off-set to the observation of $z=0.05$ (orange spectra) to show them above the observation of $z=0.17$ (green spectra). In the second column from the left, we show the measured line of sight velocity as described by Equation\,\ref{eq:los_vel_spc}. In the third column from the left, we show the observed spectra convolved by the SDSS mean instrumental broadening of 61\,km\,s$^{-1}$. On the rightmost column, we show this signal binned to the detector resolution of the SDSS with a bin-width of 69\,km\,s$^{-1}$. We fit a single and a double Gaussian function to the observations presented by yellow and black lines, respectively. For the double Gaussian function we show its blueshifted (resp. redshifted) component by dotted blue (resp. dashed red) lines. The different rows show different values of inclination as indicated in the titles.}
\label{fig:Sa_dps}%
\end{figure*}
Figure~\ref{fig:equi_vel} shows the mass surface density of the gas disc of this Sa galaxy for two different disc inclinations. Iso-velocity curves with values spaced by 30~km/s are over-plotted, starting at 0~km/s on the minor axes. The line-of-sight velocity $V$ is such that:
\begin{equation}    \label{eq:vlos}
    V = V_{\rm rot} \cos \phi \sin i,
\end{equation}
with $V_{\rm rot}$ the rotation velocity (obtained from the rotation curve, assuming a zero gas velocity dispersion), $\phi$ is the azimutal angle in the disc plane ($\phi = 0 \, [\pi]$ on the major axis), and $i$ is the inclination of the disc with respect to the line-of-sight ($i=0$ for a face-on disc). The larger the inclination of the disc, the larger the amplitude in line-of-sight velocity: for $i=30^\circ$ the iso-velocity contours have extreme values of -120 and 120~km/s (closed iso-contours near the major axis) while for $i=60^\circ$, the smallest and largest values are -210 and 210~km/s. The distribution of line-of-sight velocities is thus wider for larger inclinations, and the number of iso-velocity curves encompassed by a given fibre size is larger, as can be seen from the two represented fibres, of diameters 3 and 10~kpc. 

The fraction of gas observed with a line-of-sight velocity $V$, i.e. the line-of-sight-velocity spectrum, can be computed following \citet{1997A&A...323..727W}, as:
\begin{equation}\label{eq:los_vel_spc}
\dfrac{\mathrm{d} M}{\mathrm{d} v} (V)= \int_0^{R_{ \rm max}} \dfrac{\Sigma_{\rm gas}(R) R \mathrm{d} R}{ V_{\rm rot} (R) \sqrt{1-\left(\dfrac{V}{V_{\rm rot}(R) \sin i}\right)^2} \sin i}
\end{equation}
where the integration goes from $R=0$ to a maximum galactocentric radius $R_{ \rm max}$ (corresponding to the simulated SDSS fibre, for example), and $\Sigma_{\rm gas}(R)$ is the gas surface density. In particular, a double-horn profile can be found for a constant $V_{\rm rot}$ (see \citet{1997A&A...323..727W}). However, the formula is only approximate when applied for a radius R$_{ \rm max}$ smaller than the disc size, with an error increasing with inclination. We thus use simulated models of gas discs with Miyamoto-Nagai density profiles, setting the rotation velocity from the modelled rotation curve, and we measure the line-of-sight velocity of gas inside the different fibres for different inclinations.

We simulate the detection of double peaks for this Sa galaxy with the fibres of diameters 3 and 10~kpc for four different inclinations of the disc in Fig.~\ref{fig:Sa_dps}. The spectra obtained with Eq.~\ref{eq:los_vel_spc} and shown on the second column from the left are, as explained in Sect.~\ref{ssect:synthetic_dp}, convolved with the mean instrumental broadening of $61\,{\rm km\,s^{-1}}$ from the SDSS spectral detector (the result of this convoion is shown in the third column from the left), and then binned with the SDSS bin-width of $\delta v=69\,{\rm km\,s^{-1}}$. The spectra are broader for higher inclinations because of the $\sin i $ term in Eq.~\ref{eq:vlos}. For fixed gas density-profile and rotation curve, the shape of the spectra depends on the fibre size. For the small fibre, encompassing the beginning of the rise of the rotation curve, the spectra are single-peaked. However, a "double-horn" structure, with a central dip and sharp vertical limits at terminal velocities, appears for the larger fibre size. When viewing the disc edge-on, the double-horn shape changes to a box-like shape. This is due to the fraction of the disc moving perpendicular to the observer and which is only covered at an edge-on perspective. The instrumental broadening significantly alters the emission-line shape: e.g. the maxima of a horn are made closer to the centre of the spectrum, making the spectra single-peaked for low inclinations, and the steepness of the edges is reduced. The result of the binning is shown on the rightmost column, with both a Gaussian fit and a double-Gaussian fit. For the latter fit, the two components are shown in blue and red. Using the three criteria of Sect.~\ref{ssect:synthetic_dp}, a double-peak is identified only for $z=0.17$ (10~kpc diameter fibre) for an inclination of $90^{\circ}$. The difference of velocities of the two peaks $\Delta v_{\rm DP}$ is too small in the other cases for a double-peak to be identified according to our criteria.

Depending on the amount of dust within the line of sight, the signal of each gas particle decreases. As shown by \citet{2000MNRAS.313..153B} and \citet{2000MNRAS.318..798B}, this can cause a significant decrease in the intensity at 0~km~s$^{-1}$ and alter the emission line shape. This effect would favour a DP structure and might lead to a higher DP detection rate. However, the inclusion of this effect is not straightforward. The estimation of dust extinction is strongly depending on the wavelength \citep{1999PASP..111...63F} and factors like the dust-to-gas mass ratio \citep{1978ApJ...224..132B} and the metallicity \citep{2020ARA&A..58..529S}.
In practice this means that for the simple galaxy models chosen in this work it would be difficult to select a certain set of extinction models. In addition, since we are interested in the qualitative question of how different mechanisms can cause DP signatures, we will not include dust extinction in this work.

\subsubsection{Effect of total mass concentration on the emission-lines}
\begin{figure*}
\centering 
\includegraphics[width=0.98\textwidth]{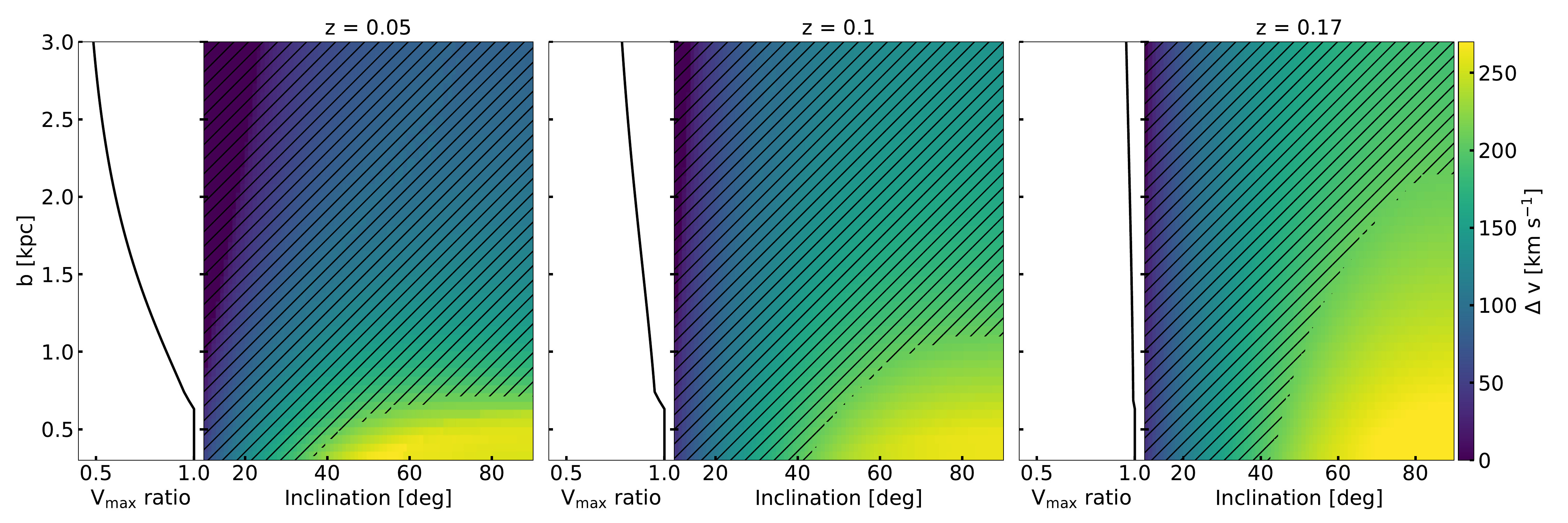}
\caption{Scans of DP detections for different inclinations and bulge concentrations of a modelled galaxy. Following Eq.\,\ref{eq:los_vel_spc}, we computed the line-of-sight velocity profile as a function of inclination and the characteristic radius of the bulge $b$. We perform the DP selection procedure described in Sect.\,\ref{ssect:synthetic_dp}. We show from left to right the results for a redshift $z=0.05$, $z=0.1$ and $z=0.17$, respectively. In each of the three panels, we show the $\Delta v$ resulting from the double-Gaussian fit with the colour coding. We mark the parameter combinations where we do not detect a DP profile with black hatches. On the left side of each panel, we show the ratio between the maximal velocity value inside the spectroscopic fibre and the maximal velocity found in the entire rotation curve.}
\label{fig:scan_bulge_dps}%
\end{figure*}

For a given (non-zero) disc inclination and a given fibre size, the detection of a double-peak is favoured by a combination of a gas density profile and a rotation curve such that more gas is probed at large line-of-sight-velocities than at small velocities (corresponding to gas on the minor axis of the disc). In order to show the effect of the shape of the rotation curve, which depends on the total mass concentration, we now keep a constant gas density profile, constant stellar disc and dark-matter halo profiles, but change the steepness of rising of the rotation curve by varying the concentration of the stellar bulge. The effect of this change is visible on Fig.~\ref{fig:rot_curves}, in which the scale length of the bulge spans from 0.3~kpc to 3~kpc (decreasing the mass concentration of the bulge and hence also of the galaxy). The rotation curve rises monotonously in the first 5~kpc for large scale lengths (low mass concentrations) while it peaks very near the centre of the galaxy for small scale lengths (high mass concentrations).   

The difference of velocity of the two peaks obtained by the fitting procedure of Sect.~\ref{ssect:synthetic_dp}, $\Delta v_{\rm DP}$, is represented (colour-coded) for different bulge scale-lengths and disc inclinations on the three panels of Fig.~\ref{fig:scan_bulge_dps}, with one panel per redshift (fibre size). The part of the rotation curves encompassed by the fibres can be seen on Fig.~\ref{fig:rot_curves}, while on the sub-panels of Fig.~\ref{fig:scan_bulge_dps} at the left of each main panel, we represent the ratio of the maximal velocity value inside the fibre to the maximal velocity in the rotation curve. Double peaks are identified with our criteria in the non-hatched regions of the panels of Fig.~\ref{fig:scan_bulge_dps}. At a given bulge scale-length, $\Delta v_{\rm DP}$ increases with inclination because of the broadening of the velocity distribution. At fixed inclination, $\Delta v_{\rm DP}$ increases with the concentration of the bulge (with decreasing bulge scale-length), with a steepness of the increase more pronounced for a small fibre. For the most mass concentrated galaxy models with a high rotation curve peak close to the centre of the galaxy, a double-peak is thus detected at small inclinations $40^{\circ}$ for all redshifts. At a given mass concentration (bulge scale-length), the threshold inclination for the double-peak detection generally increases with decreasing redshift (fibre size), with no detection for scale-lengths $> 0.7$~kpc for the smallest redshift and for scale-lengths $> 1.1$~kpc (resp. $> 2.7$~kpc) for the intermediate (resp. highest) redshift. 

\subsubsection{Effect of gas-disc concentration on the emission-lines}

\begin{figure*}
\centering 
\includegraphics[width=0.98\textwidth]{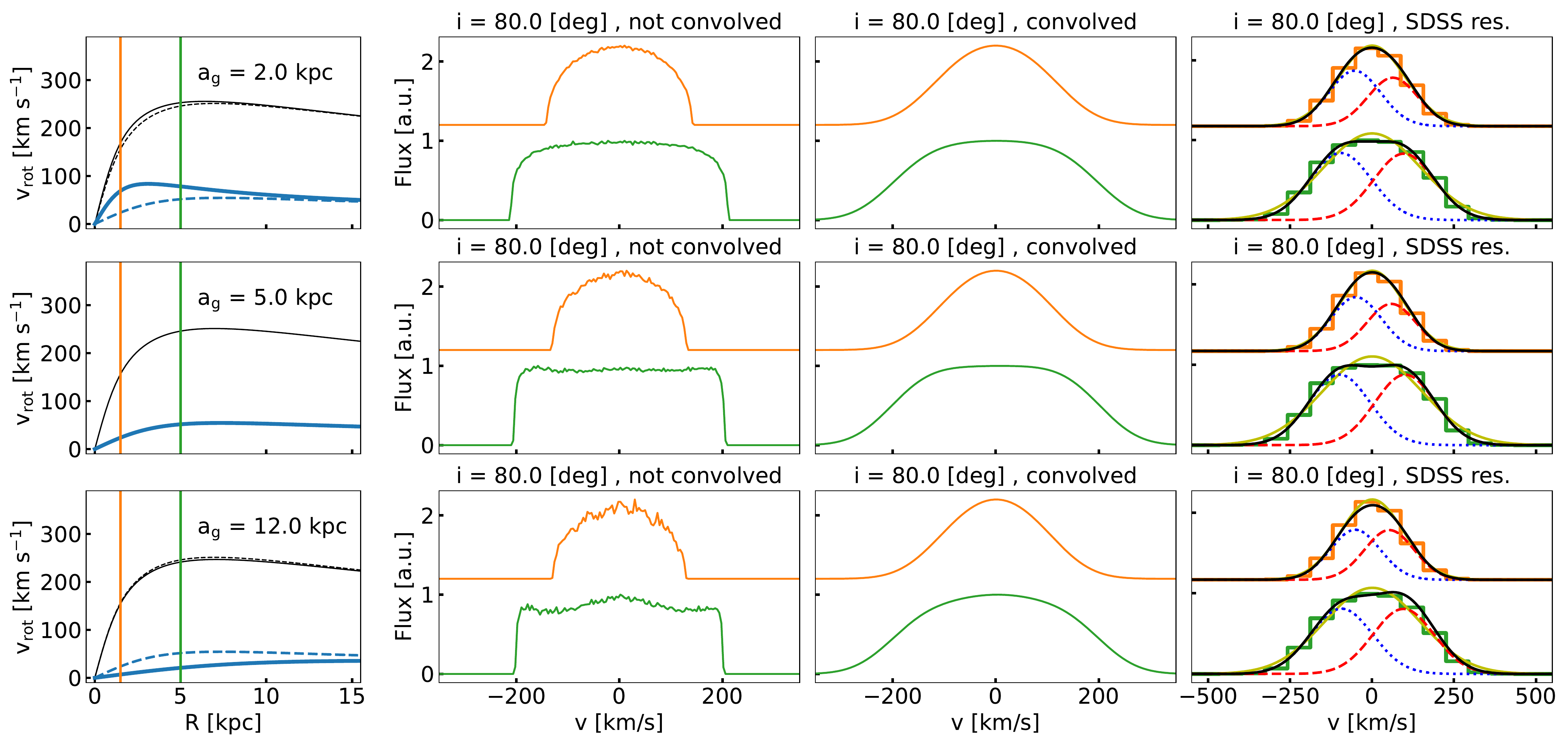}
\caption{Emission-line profiles of gaseous-disc models with different scale-lengths, with a disc inclination of $80^{\circ}$. On the left columns, we show the total rotation curves (black) and gas disc contribution (blue) in solid lines and the reference $a_{\rm gas}=5$~kpc (middle panel) in dashed lines for the top and bottom plots. For a description of the other panels, see the caption of Fig.~\ref{fig:Sa_dps}.}
\label{fig:Sa_dps_gas_prof}%
\end{figure*}

The shape of the spectra and the double-peak detection depend on the gas density-profile, which we qualitatively show on Fig.~\ref{fig:Sa_dps_gas_prof}, varying only the scale-length of the gas density profile. Because of the relative small mass of the gas component in this galaxy model with respect to the other components, changing the gas profile concentration alters very little the total rotation curve, as can be seen on the left column of the figure. 
For a less concentrated profile (a larger scale-length), the spectra indicate steeper horn features but also a higher intensity in the centre since at an inclination of $80^{\circ}$ more gas is probed close to a zero line-of-sight velocity. 
Using the three criteria of Sect.~\ref{ssect:synthetic_dp}, a DP is identified for $z=0.17$ (10~kpc diameter fibre) for scale lengths of $5$~kpc (Sa fiducial model), while the gas of the profile with a scale length of $2$~kpc is too concentrated for a DP detection. At a scale length of $12$~kpc, we do not detect a DP as the concentration at 0\,km\,s$^{-1}$ is leading to a more single-Gaussian shape. 
With the smallest fibre size, we do not detect any DP signatures. However, we observe the largest $\Delta v$ value for the smallest scale-length of $2$~kpc with $\Delta v = 117$\,km\,s$^{-1}$. For a scale length of $5$~kpc (resp. $12$~kpc), we find $\Delta v = 106$\,km\,s$^{-1}$ (resp. $\Delta v = 103$\,km\,s$^{-1}$). This is not a strong trend but it shows that for higher central gas concentrations we can see a larger contribution of the rotation in small fibres.

\subsection{N-body simulations of isolated disc galaxies}\label{ssect:n_body_sim}
The kinematic signature of emission lines is a direct probe of the gas distribution inside the spectroscopic observation area. In reality, gas is found in clumps, discs, rings, spiral arms and bars. Such structures deviate significantly from a model of an axisymmetric disc with a simple density profile such as described in Sect.\,\ref{ssect:axi_sym_model}. In order to explore how DP signatures can be found in more realistic isolated galaxies, we will here analyse simulated isolated disc galaxies. We make use of the simulations database {\sc GalMer}, which is described in detail in \citet{2010A&A...518A..61C}. This database is designed to systematically explore galaxy mergers with various initial orbital parameters, galaxy inclinations and galaxy types. To understand how galaxies evolve in isolation in comparison to the interactions, this database provides isolated galaxy simulations for each morphological type. The reading and analysis of the outputs of the simulations is based on the visualisation software \textsc{GalaXimView}\footnote{{\url{https://vm-weblerma.obspm.fr/~ahalle/galaximview/}}}.

\subsubsection{Simulation design}\label{sssect:sim_design}
We here explore the evolution of isolated Sa and Sb galaxies. In Sect.\,\ref{sect:merger}, we will further explore major-merger (giant + giant) and minor-merger (giant + dwarf) systems. The simulated isolated galaxies are giant galaxies and we thus refer to them as gSa and gSb.   
DP emission lines are mostly found in S0 and spiral galaxies of the type Sa and Sb. In the {\sc GalMer} database, S0 galaxies are designed without a gaseous disc since this galaxy type is usually observed with an exhausted gas content \citep[e.g.][]{2015ARA&A..53...51S}. 
The gSa and gSb galaxies considered here consist of rotating gas and stellar discs, a non-rotating stellar bulge and a non-rotating dark-matter halo. The initial conditions of simulations are modelled with the same density profiles as the axisymmetric models described in Sect.\,\ref{ssect:axi_sym_model}: disc components are described by a Miyamoto-Nagei density profile and the stellar bulge and dark-matter halo by a Plummer density profile. Velocities are set by the method of \citet{1993ApJS...86..389H}. The discs components have initial Toomre parameters of $Q=1.2$.

The simulation code is described in detail in \citet{2007A&A...468...61D}. It uses a Tree algorithm for the computation of the gravitational forces \citep{1986Natur.324..446B} and smoothed particle hydrodynamics \citep{1977AJ.....82.1013L,1982JCoPh..46..429G} for the gas with individual smoothing lengths. The gas is considered as isothermal with a temperature ${\rm T_{gas} = 10^4 K}$. To emulate star formation, hybrid particles, corresponding initially to pure gas particles with a stellar fraction of 0, are gradually changed into stellar particles following a star formation law described in \citet{1994ApJ...437..611M}. Once the gas fraction drops below 5\,\%, a hybrid particle is converted into a stellar particle. During the star-formation process, the total mass of the hybrid particle is constant. There is no feedback from AGN, but there is stellar mass loss, and energy re-injected in the ISM by supernovae, cf \cite{2010A&A...518A..61C}.
Time integration is performed with a leapfrog integrator with a time-step $\Delta t = 5 \times 10^5 {\rm yr}$ and snapshots are output every $5 \times 10^7 {\rm yr}$. The simulations are carried out for a time-span of 3 or 3.5\,Gyr.
The initial parameters for gSa and gSb galaxies are given in the Appendix\,\ref{app:init_param}, Table\,\ref{table:init_params}. Isolated galaxies are simulated with a total number of 480\,000 particles and a softening length of $\epsilon=200$\,pc. The same softening length is used for giant-dwarf interaction simulations while a softening length of $\epsilon=280$\,pc is used for giant-giant interactions (See Sect.\,\ref{sect:merger}). 

\subsubsection{Characterisation of the structure of the galaxies }\label{ssect:characteristic_params}
In order to conduct a systematic analysis of simulated galaxies, we compute at each simulation step the following characteristic values: the position and the velocity of the centre of baryonic mass (COM), the half-mass radius $r_{1/2}$, and the spin vector of the stellar particles. 
We calculate the COM from the baryonic particles (gas + stars). Therefore, we compute a 3D histogram with a bin-width of 1\,kpc and select the bin containing the highest mass. We then calculate the position and the velocity of the COM of the particles inside this bin. 
For each COM, we calculate the $r_{1/2}$, describing the radius containing half of the baryonic mass of a galaxy.  
The spin vector of each galaxy is estimated by calculating the angular-momentum vector of the stellar particles which are outside the $r_{1/2}$ but within a radius $<15$\,kpc. In bulge-dominated galaxies with a large central velocity dispersion, the spin vector, computed with all particles, would not weigh sufficiently the rotation of the outer disc. Hence, a spin vector, calculated only with the outer particles, provides a better approximation of the disc orientation. 
As it will be discussed in Sect.\,\ref{sect:merger}, during a violent merger with complex geometry and kinematics, this vector does not have any meaningful direction and will only be considered as a point of reference.  
In the following, spectroscopic observations are computed from an observer perspective, orientated with a polar angle $\theta$ and an azimuthal angle $\phi$ defined with respect to the spin vector and to a reference vector in the plane orthogonal to it for $\phi$. When the spin vector truly defines a disc plane, $\theta = 0^{\circ}$ (resp $\theta = 90^{\circ}$) corresponds to a face-on (resp. edge-on) observation. The inclination angle $i$ is thus $i=\theta$ for $\theta \in [0, 90^{\circ}]$ and $180^{\circ} - \theta$ for $\theta \in [90^{\circ}, 180^{\circ}]$.     

\subsection{Double-peak signatures from bars}\label{ssect:bars}
The initial conditions of the simulated galaxies are computed with the exact same models as discussed in Sect.\,\ref{ssect:axi_sym_model}. However, one important aspect is a velocity dispersion which is not included in the line-of-sight velocity distribution with the previous models. Comparisons between a simulated gSa galaxy and an axisymmetric model lead to the same DP detection dependencies. For low inclinations (nearly face-on), we find larger emission-line profiles than in the axisymmetric model, which is due to the contribution of the velocity dispersion. 
The additional velocity dispersion broadens the emission-line profile and we can therefore detect a DP signature at lower inclinations. As visualised in Fig.\,\ref{fig:scan_bulge_dps}, we detect a DP signature for inclinations larger than $70^{\circ}$ using the axisymmetric model with a parametrisation of the fiducial gSa galaxy. For the initial conditions of a simulated gSa galaxy, we detect a DP signature for inclinations larger than $50^{\circ}$, due to the contribution of the velocity dispersion.

\begin{figure}
\centering 
\includegraphics[width=0.48\textwidth]{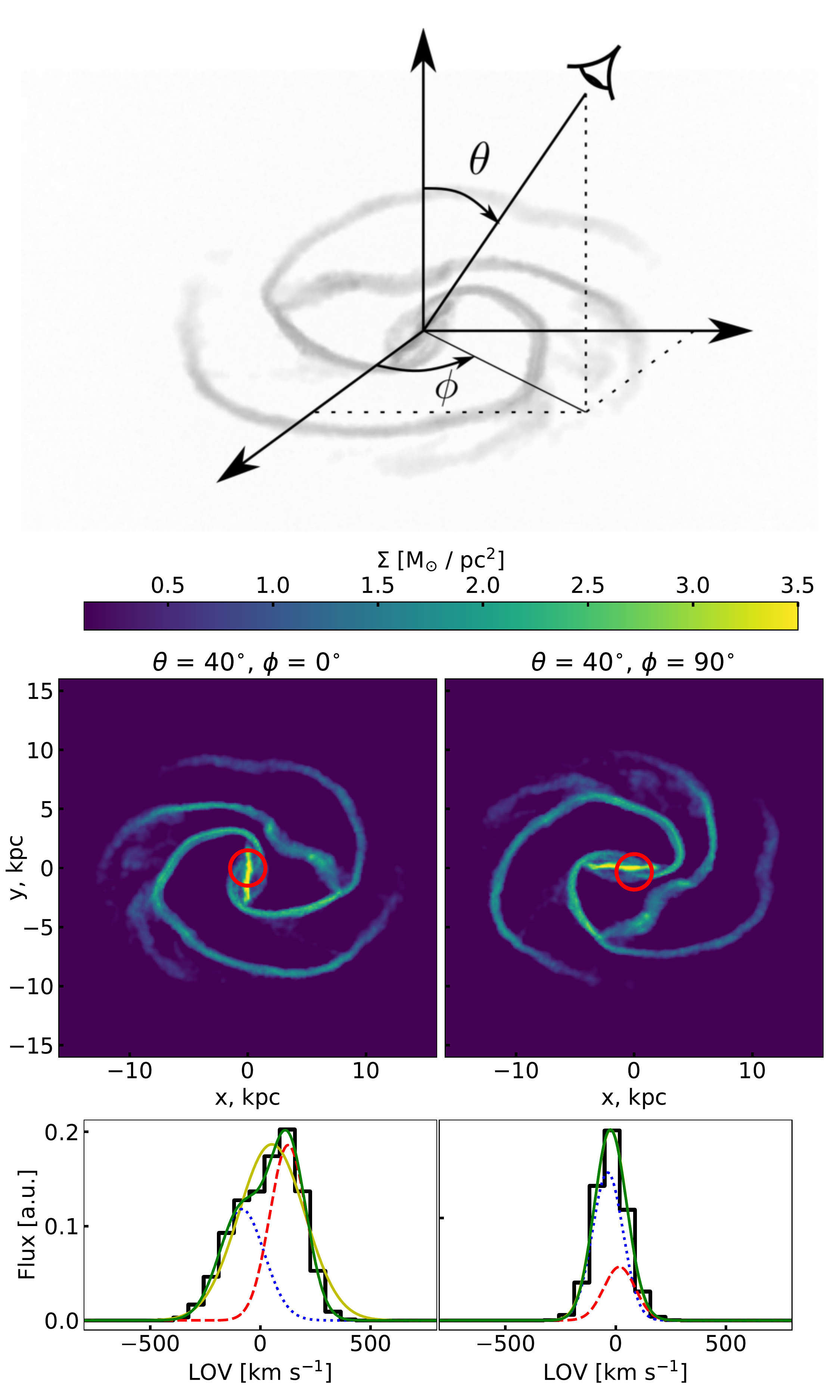}
\caption{Observation of an isolated barred galaxy. On the top panel, we show the gas distribution in the 3D space and define the definition of the observation angles $\phi$ and $\theta$. On the middle panel, two 2D projections are shown for an inclination of $\theta=40^{\circ}$. On the left (resp. right ) we show an azimuth of $\phi=0^{\circ}$ (resp. $\phi=90^{\circ}$) which corresponds to an observation parallel (resp. perpendicular) to the bar. With red circles, we mark a $3^{\prime\prime}$ spectral fibre observation situated at a redshift of $z=0.05$. On the bottom panels, we show the gas emission line line-of-sight distribution inside the fibre. We fit a double and single Gaussian function to the emission lines.}
\label{fig:bar_alignment}%
\end{figure}
The simulated galaxies undergo a rapid evolution in the first 0.5-1\,Gyr. Gas condensates into thin and dense structures and clumps, spiral arms and a stellar bar are formed. These features however vanish after at least 1\,Gyr. We observe a homogenisation of the disc with no arm structure while most of the gas has fallen into the centre. This high central gas concentration is then dominated by velocity dispersion and no DP emission-line structure can be observed any more. This likely unrealistic evolution stage is favoured by the low supernovae feedback and absence of AGN feedback in the simulations.
From observations we know that about two thirds of disc galaxies are barred \citep[e.g.][]{2000AJ....119..536E,2007ApJ...657..790M}. However, this does not imply that bars have a long life time. In fact, bars can be weakened or destroyed \citep{2005MNRAS.364L..18B}, but with a high gas fraction they can be re-formed \citep{2002A&A...392...83B}. Relying on cosmological simulations, the bar fraction is expected to be constant at about 66\% for massive spiral galaxies (${\rm M_* \geq 10^{10.6}M_{\odot}}$) over a redshift range of $z=0-1$  \citep{2020ApJ...904..170Z}.

Gas clumps, spiral arm structures and turbulence in the simulations lead to some minor fluctuations of the DP detection. A stellar bar, however, is significantly changing the DP detection: we find strong $\Delta v$ values of more than 300\,km\,s$^{-1}$ when observing parallel to the bar at an inclination of $\theta=60^{\circ}$. 
Observations of a gSa galaxy with a characteristic bar structure is shown in Fig.\,\ref{fig:bar_alignment}, after an evolution of 250\,Myr from the initial axisymmetric condition. 
We define the observation angles in the top and show the 2D-projection of the observed gas in the middle panels: on the left, the disc is seen parallel to the bar and on the right, perpendicular to the bar. On the bottom panels, we show the spectroscopic observation of the gas for the two cases. We find a strong DP feature in the observation taken parallel to the bar but no DP signature in the one observed perpendicular to the bar. This is due to the fact that when observing perpendicular to the bar, the majority of the gas is moving also perpendicular to the line-of sight. Hence, we do not probe a large velocity gradient. In comparison to that, when observing parallel to the bar, we measure gas moving alongside the line of sight due to its streaming motion along the bar. 

\begin{figure}
\centering 
\includegraphics[width=0.48\textwidth]{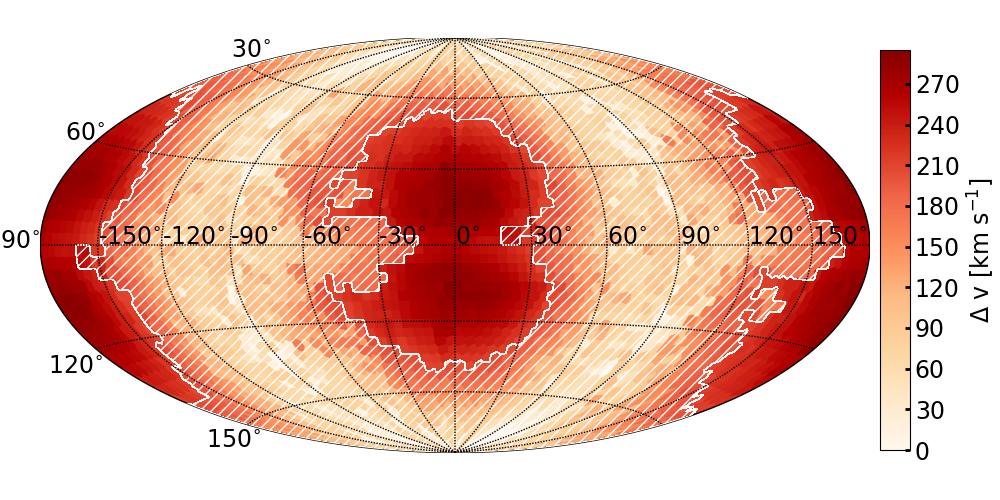}
\caption{$\Delta v$ values measured at different observation angles with a double Gaussian fit at a redshift $z=0.05$. Each individual measurement covers a solid angle of 0.013\,sr and the colour code indicates the measured $\Delta v$ value. We choose the Hammer-projection to represent the observation points on the surface. The longitudes represent the azimuth angle $\phi$ which is measuring the observation angle relative to the central bar of the galaxy.  $\phi=0^{\circ}$ and $\phi=\pm 180^{\circ}$ correspond to an observation parallel to the bar and an azimuth angle $\phi=\pm 90^{\circ}$ to an observation perpendicular to the bar. The latitudes represent the inclination of the observer. At an inclination of $\theta=90^{\circ}$ the galaxy is observed edge-on whereas at $\theta=0^{\circ}$ and $\theta=180^{\circ}$ one sees the galaxy face-on.}
\label{fig:isolated_scan}%
\end{figure}
In order to compute from which observation angles one can find a DP signature, we systematically place the observer on a sphere around the galaxy with the COM as its centre. We choose a uniform sampling of the sphere so that each observation covers a solid angle of 0.013\,sr. 
In Fig.\,\ref{fig:isolated_scan}, we show a scan of all observation angles observed at $z=0.05$ for the gSa galaxy exhibiting a bar which is visualised in Fig.\,\ref{fig:bar_alignment}.
We indicate the $\Delta v$ computed from the double Gaussian fit with a colour code and mark the angular positions that do not exhibit a DP signature with white hatches.  
We show the full map (here and in other figures) but note that in the absence of any attenuation, the map contains redundant information: the value at $\theta$ and $\phi$ is the same as the value at $180^{\circ} - \theta$ and $\phi + 180^{\circ}$ (modulo $360^{\circ}$). 
If DP signatures originated from uniform rotation, a DP would be observed at all azimuth angles with a strong inclination of $60^{\circ} < \theta < 120^{\circ}$ as we found in Sect.\,\ref{ssect:axi_sym_model}. However, this is not the case: we see a strong DP signature when observing parallel to the bar ($\phi \sim 0^{\circ}$ and $\phi \sim \pm 180^{\circ}$) and single-peak signatures when observed perpendicular to it. 
Furthermore, we do not see the highest $\Delta v$ values when observing fully edge-on ($\theta = 90^{\circ}$) but at a lower inclination of $\theta \sim 75^{\circ}$. This is due to the fact that when observing fully edge-on along the bar direction, the spectroscopic measurement probes as well gas, at the ends of the bar or elsewhere in the disc, moving perpendicular to the observer and contributing to the line-of-sight velocity distribution at $v=0 \, {\rm km\,s^{-1}}$. This makes the two Gaussian functions of the double Gaussian fit shift closer together and the $\Delta v$ become smaller. In contrast to that, when observing at a smaller inclination, the observation fibre of 3\,kpc in diameter (seen at a redshift of $z=0.05$) will mostly probe gas with a motion along the bar direction. This gas moves at the highest velocity parallel to the line-of-sight and only a small contribution of gas moving perpendicular is measured. This leads to a strong DP feature. 

\begin{figure}
\centering 
\includegraphics[width=0.48\textwidth]{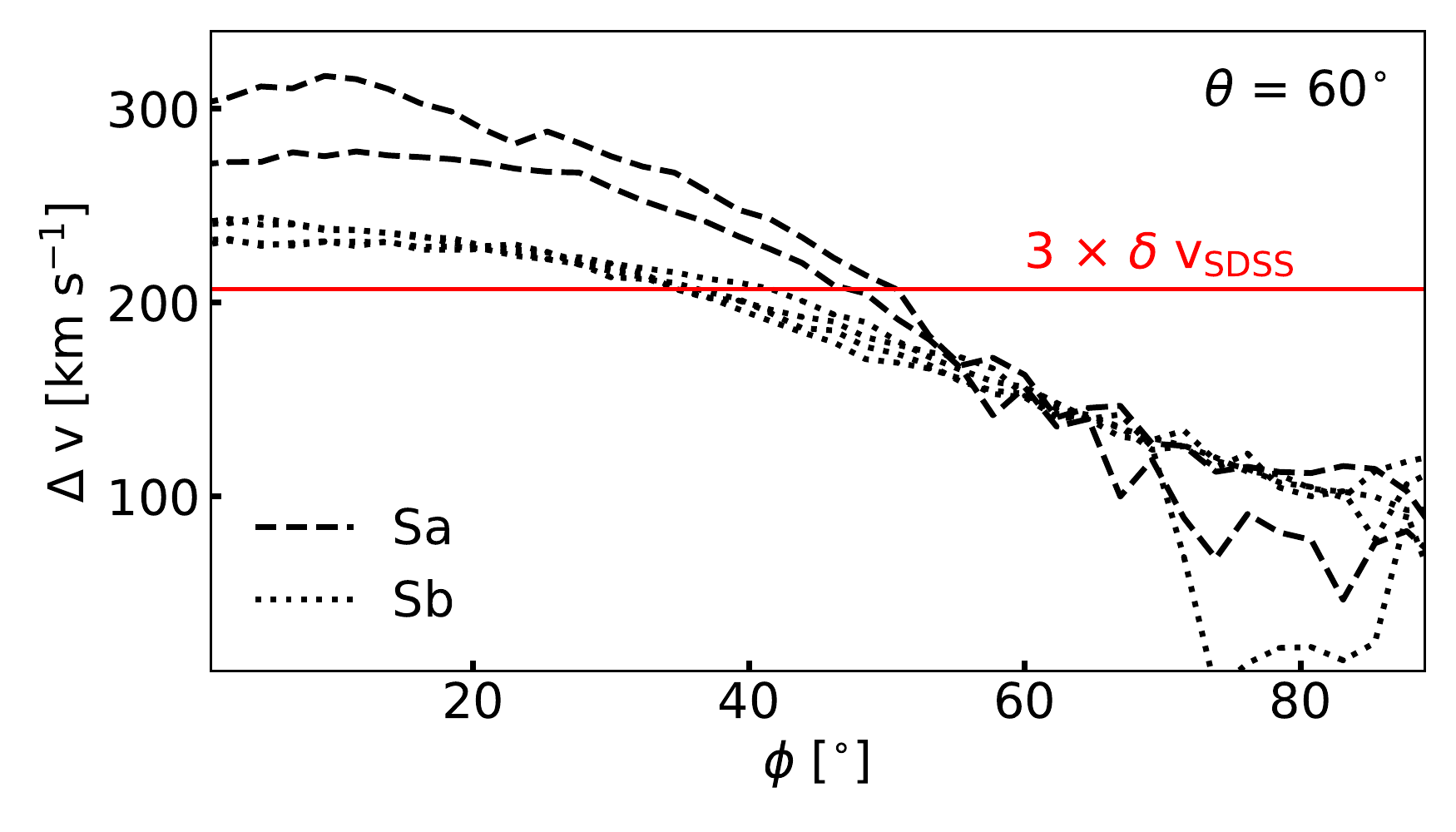}
\caption{$\Delta v$ values from a double Gaussian fit at different azimuth angles $\phi$. We define $\phi$ as the azimuth angle with respect to the bar as visualised in Fig.\,\ref{fig:bar_alignment}. We computed spectroscopic observations at a fixed inclination of $\theta=60^{\circ}$ and at a redshift of $z=0.05$}. We show all snapshots of gSa and gSb simulations, which indicate a bar. With a red line we show the value of three times the bin-width of the SDSS. A $\Delta v$ larger than this value is one criteria for a DP detection (see Sect.\,\ref{ssect:synthetic_dp}).
\label{fig:bar_scan}%
\end{figure}
This effect can be seen on further galaxy examples. In Fig.\,\ref{fig:bar_scan}, we have included all snapshots of gSa and gSb simulations which have a bar. We have determined the $\Delta v$ value for a constant inclination of $\theta = 60^{\circ}$ and for azimuth values between $\phi = 0^{\circ}$ (parallel to the bar) and $\phi = 90^{\circ}$ (perpendicular to the bar). The spectroscopic observations are evaluated within a spectroscopic fibre of a diameter of 3\,kpc, corresponding to a SDSS fibre at redshift $z=0.05$.
For observations with an angle $\phi$ up to $\sim 40^{\circ}$, we find $\Delta v$ values exceeding three times the SDSS bin-width ($3\times \delta v_{\rm sdss}$), the definition threshold for a DP signature (see Sect.\,\ref{ssect:synthetic_dp}). We find higher $\Delta v$ values for snapshots of the gSa galaxy in comparison to the gSb galaxy, because of the more massive stellar bulge of the gSa galaxy, resulting in a deeper gravitational potential and thus in faster rotation in the centre. For all observations, the $\Delta v$ value drops below the threshold of $3\times \delta v_{\rm sdss}$ when observing perpendicular to the bar.
This means that gas motion created by bars can indeed be at the origin of a strong DP feature. 
Furthermore, as a bar seen parallel to the line-of-sight is difficult to identify as such in galaxy images, the fraction of DP galaxies in observational studies may show a deficit of bars while bars are in fact the origin of a part of the double peaks. 

We can compute the DP fraction ${\rm f_{DP}}$ as the fraction of directions from which a DP signature is observed. When observing at a redshift $z=0.05$, the mean DP fraction is ${\rm f_{DP}} = 0.24$. This fraction drops to a mean value of ${\rm f_{DP}} = 0.2$ at redshift $z=0.1$. 
As the fibre covers a larger part of the galaxy, more gas at lower velocities is included in the line-of-sight measurements,
diluting the DP signature. 
However, at $z=0.17$ we detect a mean DP fraction ${\rm f_{DP}} = 0.24$. This DP feature is partly originating from the bar and partly from a rotating disc. The latter effect becomes significant only at higher redshift as a larger part of the rotating disc is included in the line-of-sight velocity measurement.

\section{Mergers and post-mergers}\label{sect:merger}
In the previous section, we showed that a DP signature can be the result of a rotating disc or a bar. However, in the course of a galaxy merger, two gas components can fall into the gravitational potential well of the interacting system with different line-of-sight velocities. This, in turn, can be observed as DP emission lines in a central spectroscopic observation. Late stages of post-coalescence major mergers are known to mostly form elliptical galaxies \citep[e.g.][]{2002NewA....7..155S}.
However, the expelled gas during a merger can be re-accreted and form a disc \citep[e.g.][]{2002MNRAS.333..481B, 2006ApJ...645..986R, 2008MNRAS.391.1137L, 2009A&A...493..899P}. 
Merger events can cause a contraction of a gas disc which then forms a central rotating star-formation site \citep{2014MNRAS.438.1870D}. Such a nuclear disc can have a DP emission-line signature. 
Since a single minor merger is not expected to cause radical morphological transformations, we examine, besides major mergers, also the possibility of how a minor merger can funnel gas into the central region and create a DP emission-line signature.  

In order to explore a DP signature which is related to galaxy mergers, we here explore major mergers with a mass-ratio of 1:1 (giant + giant) and minor mergers with a mass ratio of 1:10 (giant + dwarf). 
As discussed in \citetalias{2020A&A...641A.171M}, DP signatures are mostly associated with spiral galaxies of type Sa and Sb and S0 galaxies. At high redshift, it is difficult to distinguish an elliptical galaxy of e.g. Hubble type E6 from a S0 or Sa galaxy. This motivates merger scenarii leading to earlier Hubble types. 
We thus select from the {\sc GalMer} database the major-merger simulations gSa + gSa and gSb + gSb. For minor-merger simulations, we explore gSa + dSb and gSa + dSd. 
We evaluate possible DP signatures of the selected merger simulations from all directions, in the same way as in Sect.\,\ref{ssect:bars}, for all three representative SDSS spectroscopic fibre diameters at redshift $z=0.05$, $z=0.1$, and $z=0.17$ (see Sect.\,\ref{ssect:fibre_size}). 

We consider major-merger simulations between two galaxies of the same type, leading to an equal contribution of gas in the resulting system. Even though we selected dwarf galaxies for the minor-merger simulations with the highest gas fraction compared to the giant gSa galaxy, the resulting gas mass ratio is still of 1:10 to 1:5 (see Table\,\ref{table:init_params}). In order to identify two Gaussian components in an emission line as a DP signature, an amplitude ratio of at least three is necessary (See Sect.\,\ref{ssect:synthetic_dp}).
However, dwarf galaxies have a significant lower metallicity than giant galaxies \citep[][]{2004ApJ...613..898T}, which in fact leads to a stronger emission-line signal \citep[e.g.][]{2010ApJ...716.1191W, 2013ARA&A..51..207B, 2019ARA&A..57..511K}. Since we aim to clarify quantitatively how a minor merger can generate a DP signature through internal kinematic processes, we multiply the signal from the giant galaxy by a factor of 0.5. The choice of this factor is purely empirical, since it results in a DP detection with two Gaussian components inside the line-of-sight. However, if one aims to obtain a more complete picture of the contributions of different gas populations in galaxy mergers, an accurate calibration of a radiative transfer would be necessary.

\subsection{Merger simulation parameters}\label{ssect:merger_parameters}
\begin{figure}
\centering 
\includegraphics[width=0.48\textwidth]{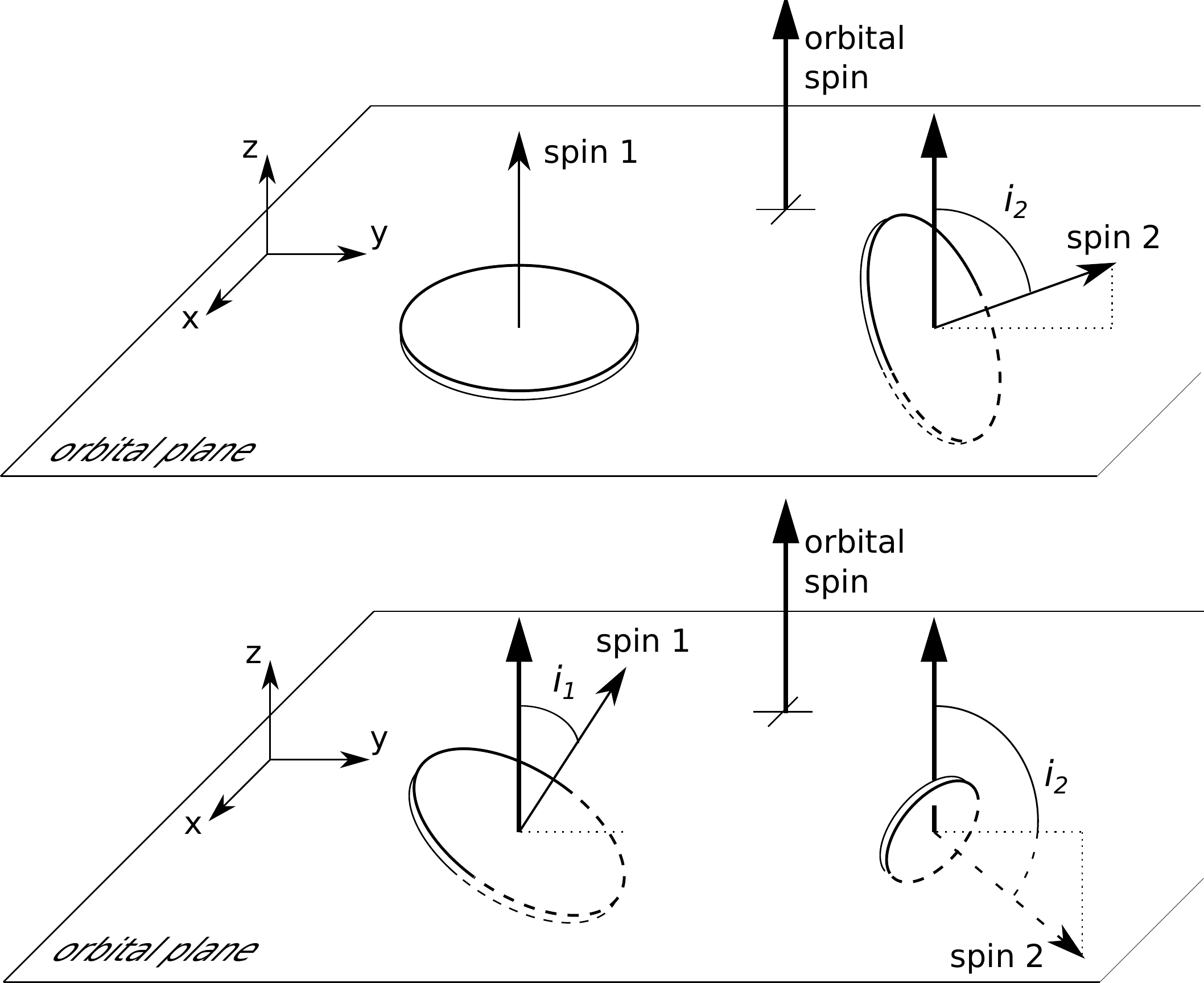}
\caption{
Top: Initial configuration of a direct-direct merger between two giant galaxies. In the retrograde-retrograde configuration, the orbital spin is flipped. The inclination $i_2$ is set to 0, 45, 75 or $90^{\circ}$. Galaxy 1 has a spin of coordinates $(0,0,1)$, and galaxy 2, $(0,\sin i_2, \cos i_2)$.
Bottom: Initial configuration of a direct-retrograde merger between a giant and a dwarf. In the retrograde-direct configuration, the orbital spin is flipped. The giant galaxy has an inclination $i_1 = 33^{\circ}$ and the dwarf galaxy, $i_2 = 130^{\circ}$. Galaxy 1 has a spin of coordinates $(0,\sin i_1, \cos i_1)$, and galaxy 2, $(0,\sin i_2, \cos i_2)$.
In both panels, dashed lines indicate that patterns are below the orbital plane or behind the discs.}
\label{fig:spins}%
\end{figure}
The {\sc GalMer} database provides major and minor galaxy merger simulations. Major mergers are simulated with a total particle number of ${\rm N_{tot}=240\,000}$ (120\,000 + 120\,000) particles, while minor mergers have ${\rm N_{tot}=528\,000}$ (480\,000 + 48\,000), i.e. 4 times more particles for a giant galaxy than in a major merger, in order to resolve the dwarf galaxy \citep{2010A&A...518A..61C}. Thus, the softening length for major mergers is $\epsilon=280$\,pc, and $\epsilon=200$\,pc for minor mergers.
The initial conditions for each galaxy are set in the same way as for isolated galaxies described in Sect.\ref{sssect:sim_design} with initial parameters for the different galaxy types summarised in Table\,\ref{table:init_params}. The two galaxies are initially set at a distance of 100\,kpc with an orbit characterised by the orbital angular momentum $\mathbf{L}$. Giant-giant mergers simulations are carried out either with a direct-direct configuration in which the spins of both galaxies have a positive projection on the orbital spin (unit vector aligned with the orbital angular momentum), or with a retrograde-retrograde configuration, where the orbital spin is flipped. For the giant-dwarf mergers, simulations are carried out either with a direct-retrograde configuration as shown in Fig.~\ref{fig:spins}, or with a retrograde-direct configuration. In the giant-giant mergers, the disc plane of one galaxy is always initially in the orbital plane while the other one has an inclination $i_2$ with respect to the plane (see Fig.~\ref{fig:spins}). The giant-dwarf configuration is more generic: both discs are inclined with respect to the orbital plane (see Fig.~\ref{fig:spins})
A detailed description of the orbital parameters for the {\sc GalMer} database is given in \citet{2010A&A...518A..61C}. In the Tables\,\ref{table:orb_param}, we summarise the orbital parameters used in this work. 

We are interested in DP emission-line signatures during mergers and after coalescence. We, therefore, sort out all fly-by simulations, in which the two galaxies only move away from each other after one single encounter, and retrograde minor mergers (with a retrograde configuration for the giant galaxy), whose final coalescence does not happen during the simulated period. 
Galaxy mergers with a retrograde orbit last longer in comparison to direct ones \citep{2012MNRAS.424.2401V, 2018A&A...614A..66S}. 
As mentioned in Sect.\,\ref{sssect:sim_design}, the simulation design of the {\sc GalMer} database does not include AGN feedback. This leads to high concentration in the very centre at the end, where almost no rotation is visible in the central gas. Depending on the merger process, the central in-fall of gas can happen with no gas being expelled. In such a situation, we do not see any DP signature and thus such scenarii are not interesting for this work and are sorted out. This happens more frequently in gSa + gSa mergers which is most probably related to the deeper gravitational potential from the final stellar bulge in comparison to gSb + gSb mergers.
These selection criteria lead us to a final simulation sample of 16 major-merger simulations (5 gSa + gSa and 11 gSb + gSb) and 11 minor-merger simulations (6 gSa + dSd and 5 gSa + dSb).

\subsection{Characterisation of the merger process and DP fraction measurement}\label{ssect:merger_characteristics}
\begin{figure*}
\centering 
\includegraphics[width=0.98\textwidth]{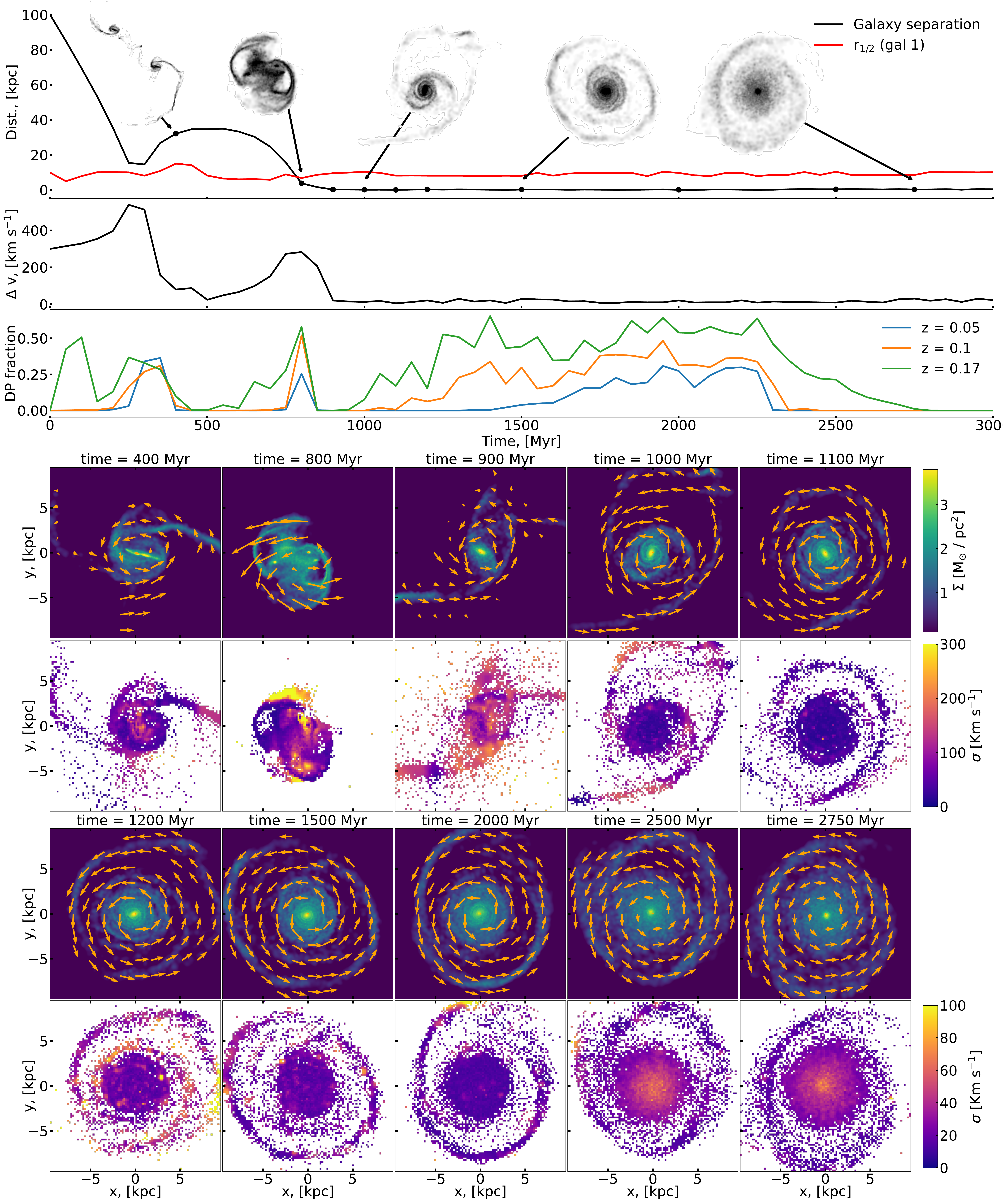}
\caption{Visualisation of a major merger process. We show characteristic parameters of the galaxy merger simulation of gSb + gSb with an orbit id 02dir and a merger inclination of $0^{\circ}$ (See Sect.\,\ref{ssect:merger_parameters} and Table\,\ref{table:orb_param}). On the top panel, we show with a black line the distance between the COM of the two galaxies which corresponds to the distance between the two galaxies. The red line represents the half mass radius r$_{1/2}$ of the first galaxy. In order to illustrate the merger process, we show snapshots of only the gas at different merger stages. Black arrows indicate the exact stage of the merger process. On the second panel, we show the velocity difference between the two COM. On the third panel, we show the DP fraction which corresponds to the fraction of observation angles from which one detects a DP. On the bottom panels, we show zoomed-in observations of the central kiloparsecs of the first galaxy, we display, on the top panels, the 2D projection of the gas surface density $\Sigma$ observed from a face-on view and the measured velocity dispersion $\sigma$ on the panels beneath. To illustrate the gas dynamics, we show arrows representing the 2D projected in-plane velocity of the particles. }
\label{fig:post_coal_sb}%
\end{figure*}
To describe the orbit of a galaxy merger, we compute the COM and r$_{1/2}$ of each individual galaxy as described in Sect.\,\ref{ssect:characteristic_params}. This allows us to compute the distance between the two galaxies at each simulation step, shown as the black line on the top panel of Fig.\,\ref{fig:post_coal_sb}. The r$_{1/2}$ value of the first galaxy is shown with a red line. In order to visualise the morphology of the gas during the merger, we show snapshots of the gas distribution of some simulation steps and use arrows to mark their position on the evolution of the simulation. In the second panel, we show the velocity difference between the two COMs. With these parameters, we can characterise the merger simulations: we clearly see the first peri-passage after about 250\,Myr. This is the point where the velocity difference between the two galaxies is the highest. The two galaxies then recede from each other until the point at about 500\,Myr where we see a maximum of their distance and a minimum of their velocity difference. The two galaxies then fall back onto each other and finally merge. We estimate some coalescence time as the time after which the distance between the two galaxies no longer exceeds the half-mass radius $r_{1/2}$ of the first galaxy. The velocity difference is also then dropping to 0. 

In order to understand at what merger stage a DP emission line signature can be observed, we scan each simulation step from all directions as described in Sect.\,\ref{ssect:bars}, with a uniform sampling of the sphere. For the major mergers, the origin of this scan is set to the COM of the galaxy whose disc is initially in the orbital plane. For the minor mergers, the origin is set to the COM of the giant galaxy. We also orientate the viewing angle with the spin of these reference galaxies.
This provides us a DP fraction at each simulation step, which we show in the third panel from the top in Fig.\,\ref{fig:post_coal_sb} and \ref{fig:post_coal_sa}. For the gSb galaxies in major-merger simulations, we do not observe any DP emission-line signature in the initial conditions. However, we find for a redshift of $z=0.17$ a DP fraction of about 0.6 for gSa galaxies which is in good agreement with the DP signatures found for an Sa galaxy with the same parameters using an axisymmetric model in Sect.\,\ref{ssect:axi_sym_model}, where we find a DP signature for inclinations larger than $\theta=55^{\circ}$, which covers about 60\,\% of a sphere.    
While the galaxies in major-merger simulations start with the initial parameters described in Sect.\,\ref{ssect:axi_sym_model}. minor-merger simulations start with already evolved galaxies. Therefore, the initial DP fraction in gSa galaxies is quite different in the initial snapshot of minor-merger simulations.

During the merger process, we always observe a peak of DP fraction during 50-100\,Myr after a peri-passage. This phenomenon of two galaxies observed in the act of merging was analysed by \citet{2021A&A...653A..47M} and will be discussed systematically in Halle et al (in prep.). 
As we know from observations, DP emission line signatures are not significantly more common in visually identified galaxy merger systems \citep{2020A&A...641A.171M}. Therefore, we here focus on DP signatures which appear in the post-coalescence phase of major and minor mergers.

\subsection{Double-peak signatures in major mergers}\label{ssect:major_merger}
Here we discuss systematically at what merger stage we can observe a DP signature. We furthermore discuss the significance of the observation angle and further discuss the morphology of the resulting galaxy. 

\subsubsection{Central discs in post major mergers}
Major mergers are known to show strong morphological perturbations during the merger. In \citet{2008MNRAS.391.1137L}, the timescale during which a merger is observable from the photometry of equal-mass galaxy mergers was estimated to be of the order of $1.1-1.9$\,Gyr. This timescale can vary due to different orbital parameters which determine when the final coalescence happens. Looking at the exemplary gSb + gSb major merger shown in Fig.\,\ref{fig:post_coal_sb} and the gSa + gSa major merger in Fig.\,\ref{fig:post_coal_sa}, there is no DP signature directly after the final coalescence. However, at about 1\,Gyr after the final coalescence, an increasing DP fraction is detected. 

On the bottom panels of Fig.\,\ref{fig:post_coal_sb} and \ref{fig:post_coal_sa}, we display 10 snapshots of the central parts of the first galaxy at different simulation steps, marked with black dots in the galaxy separation diagram. We show for each selected time the gas surface brightness and the velocity dispersion. The line of sight is parallel to the spin vector so that discs are seen face-on. Gas motion in the plane is illustrated with orange velocity arrows. 
During the simulation of the gSb + gSb merger (Fig.\,\ref{fig:post_coal_sb}), we observe a peak in DP fraction in an early phase at 400\,Myr, shortly after the first encounter. A second peak is observed at 800\,Myr, at the moment of post coalescence. In the snapshot of the central region of the first galaxy at 400\,Myr, we identify a bar structure as the origin of the increase in the DP fraction. As discussed in Sect.\,\ref{ssect:bars}, a central bar structure in the gas distribution can create strong DP signatures, especially for small spectroscopic fibre diameters. 
For the second peak in DP fraction at 800\,Myr, we can identify the two galaxies at a separation less than 4\,kpc and with a velocity difference of 300\,km\,s$^{-1}$, creating a DP signature as two gas populations with high $\Delta v$ are captured inside the spectroscopic fibres. The two galaxies are no longer moving away from each other and this moment marks the final coalescence.

Shortly after this final coalescence, the detection of DP stops abruptly. We observe in these stages a high concentration of gas in the very centre with a strong velocity dispersion which dominates in the observed region. In fact, the strong velocity dispersion is not sufficient to produce a broad emission-line profile which can be identified as a DP.
About a few 100\,Myr after the final coalescence, a gaseous central disc with a radius smaller than 5\,kpc starts to form. 
In contrast to the strong perturbations during the coalescence the gas starts to settle in the disc and the velocity dispersion decreases.
The in-falling gas originates from parts of the tidal tails which gradually fall back onto the galaxy. 
As the stellar bulge of the post-merger galaxy gradually grows, the rotation curve becomes increasingly steep in the centre. As we know from Sect.\,\ref{ssect:axi_sym_model}, a steep rotation curve is needed in order to detect a DP signature at lower redshift because e.g. at $z=0.05$, only the central 3\,kpc of the rotation curve is measured. This gradual steepening of the rotation curve explains why we start detecting a DP signature later during post-coalescence in low redshift observations than high redshift ones.

The detected DP signature eventually disappears at about 2500\,Myr. At this point, the gas contracts drastically to the very centre and the central part of the disc is dominated by random motion which can be seen as the velocity dispersion increases. 
As mentioned in Sect.\,\ref{sssect:sim_design}, this is due to low feedback efficiency. It is therefore difficult to say whether such a rapid collapse is realistic or whether a central disc can fall so quickly into the centre. Therefore, in the following, we will only consider simulation snapshots up to the moment when we also see a gas distribution that is not contracted below he resolution. 

\begin{figure*}
\centering 
\includegraphics[width=0.98\textwidth]{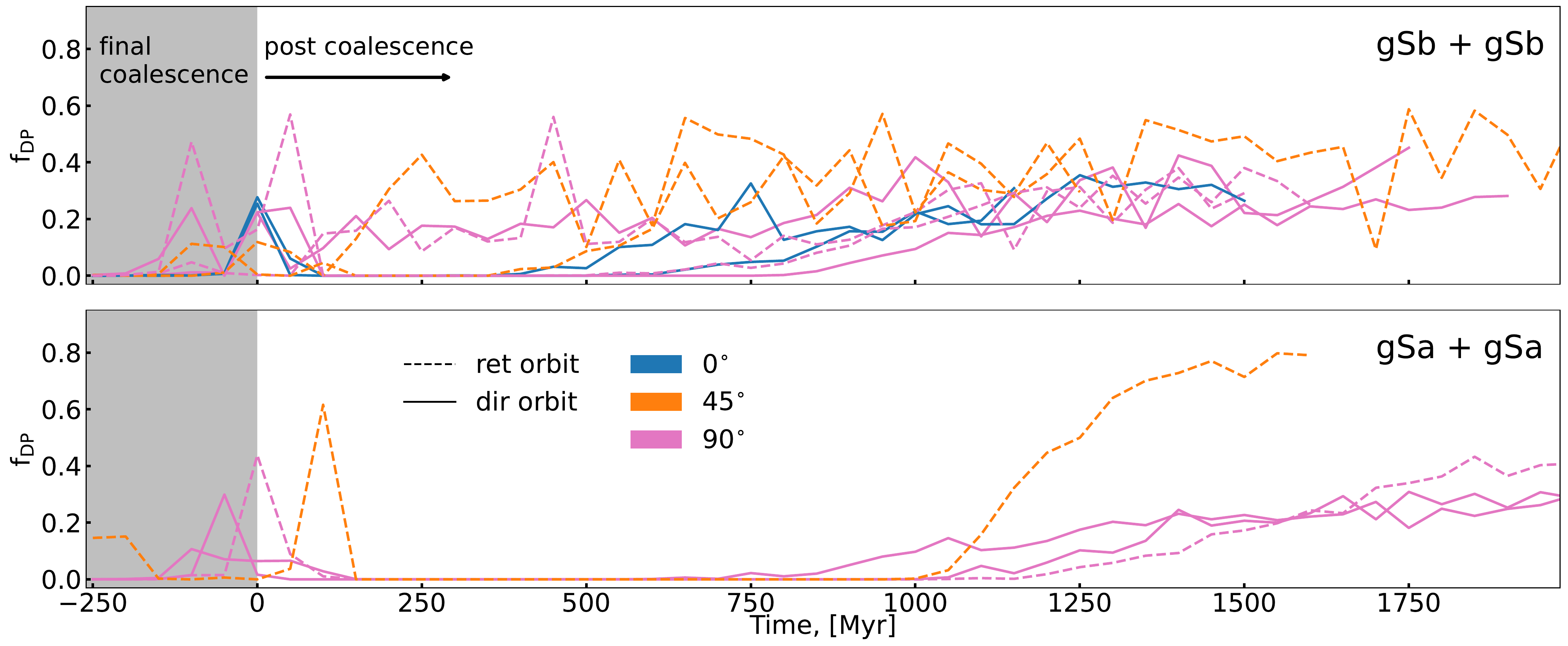}
\caption{Evolution of the DP fraction of major-merger simulations after the final coalescence observed at $z=0.05$. We identify the snapshot in each simulation where the distance between the two galaxies remains below the half mass radius of the first galaxy for the rest of the simulation. We use this snapshot as reference point and show the time starting 250\,Myr before this snapshot on the $x$-axis. On the $y$-axis, we show the DP fraction ${\rm f_{DP}}$. On the top (resp. bottom) panel, we show gSb + gSb (resp. gSa + gSa) simulations. We mark the time before the final coalescence in grey.
We mark simulations with an inclination of $0^{\circ}$, $45^{\circ}$ and $90^{\circ}$ of the second disc (see Fig.~\ref{fig:spins}) with blue, orange and pink lines, respectively. Mergers with direct (resp. retrograde) orbit are presented by solid (resp. dashed) lines.}
\label{fig:major_orbit}%
\end{figure*}
We computed the DP fraction for all selected major-merger simulations and observe a recurring pattern: strong DP detection is observed at close interactions and a gradually increasing DP fraction emerges between 500 and 1000\,Myr after the final coalescence. 
In Fig.\,\ref{fig:major_orbit}, we show the DP fraction observed at $z=0.05$ seen after the final coalescence for all selected major-merger simulations. We see, that for gSb + gSb mergers, a DP can be detected between 500 and 1000\,Myr after the final coalescence and in some cases we see a DP continuously since the final coalescence. We identify in all these cases a gaseous disc which is progressively formed from gas of the tidal tails falling into the central kiloparsecs of the galaxy.
In the post-coalescence phase of the gSa + gSa simulations, we observe an increase in DP fraction but starting 1000\,Myr after the final coalescence. 
This delay is mostly due to the fact that in comparison to gSb + gSb mergers, gSa + gSa mergers stabilise the gas due to a deeper gravitational potential and therefore, the gas takes longer to migrate towards the central region. As can be seen in the appendix in Fig.\,\ref{fig:post_coal_sa}, the central discs found in post-merger gSa + gSa galaxies are significantly smaller at a radius below 3\,kpc in comparison to the discs observed in gSb + gSb simulations.
However, this simulation stands out among other gSa + gSa simulations as it shows the smallest disc which we observed in any post-coalescence mergers. This strong concentration leads to a high DP fraction of about 0.8 at the end of the simulation. 

The post-coalescence behaviour of the DP fraction is only shown for observations at $z=0.05$ in Fig.\,\ref{fig:major_orbit}. For the observations at $z=0.1$ and $z=0.17$, we find a similar evolution of the DP fractions but higher fractions than one can see for the example shown in Fig.\,\ref{fig:post_coal_sb}. 
In Fig.\,\ref{fig:post_coal_sa}, however, we see in the late development of the central disc a larger DP fraction for the observation at $z=0.05$ which only covered the central 3\,kpc than for the other redshifts. This is due to the fact, that observations at higher redshift includes gas located outside the central disc and with more random motion, diluting the DP signal.
Considering the different merger orbits we discussed here, we do not find any dependence on the orbital geometry of the resulting DP feature.

\subsubsection{Angular dependent double-peak emission lines in the post coalescence}\label{sssect:direction_major}

\begin{figure*}
\centering 
\includegraphics[width=0.98\textwidth]{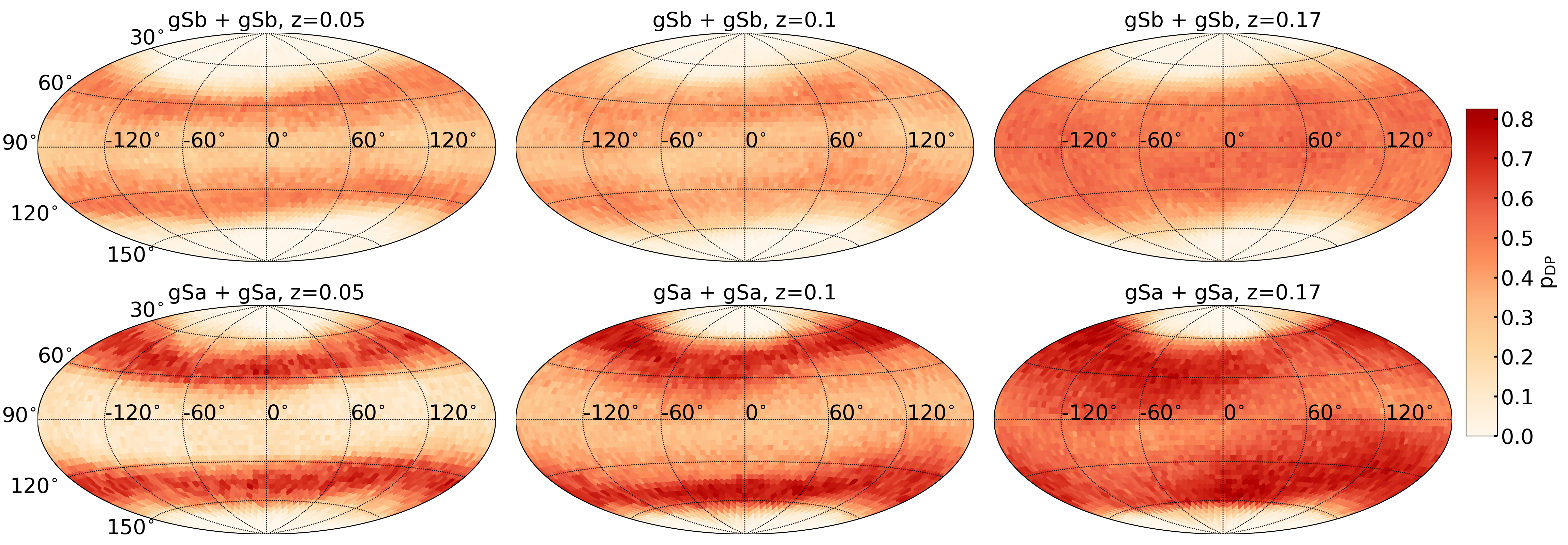}
\caption{Direction maps of DP detection probability ${\rm p_{DP}}$ in post-coalescence major-merger simulations. We present maps of ${\rm p_{DP}}$ for the three different evaluated redshifts ($z=0.05, 0.1$ and 0.17) and for the two discussed merger simulations gSb + gSb and gSa + gSa separately. We calculate the probability of each scanned direction from all simulation steps 1000\,Myr after the final coalescence that have a DP fraction $> 0.1$.}
\label{fig:dp_map}%
\end{figure*}
To visualise the observation angles from where we mostly observe a DP signature in the post-coalescence phase, we calculate an observation angular depending DP probability ${\rm p_{DP}}$. We select all simulation snapshots 1000\,Myr after the final coalescence, which show a DP fraction of at least 0.1. We then calculate a DP probability for each viewing angle as the ratio between the number of DP detections and the number of included snapshots. We do this separately for gSb + gSb and gSa + gSa simulations and further divide them into the three observed redshifts $z=0.05, 0.1$ and 0.17. This provides maps, presented in Fig.\,\ref{fig:dp_map}, indicating the most favourable observation direction for DP signatures or rule out specific observation angles.

We do not find any DP detections when observing face-on which is expected, as we observe rotating discs in all post-coalescence phases.
However, for gSa + gSa simulations we find in some cases a DP signature up to an angle of $20^{\circ}$. 
For the observations at low redshift ($z=0.05$), we see for gSb + gSb galaxies that edge-on observations are less favourable to detect a DP than observations at an inclination of about $\theta \sim 60^{\circ}$. This is due to the same reason as we discussed for DP observations in galaxies with bar signatures in Sect.\,\ref{ssect:bars}: when seen perfectly edge-on, more gas moves perpendicular to the observer. Since we observe only a small part of the velocity gradient in the central 3\,kpc at a redshift of $z=0.05$, this gas moving at a null projected velocity dominates the emission-line profile, and the DP signature produced by the rotation is not detectable anymore. However, at a smaller inclination, significantly less gas moving perpendicular to the observer contributes to the observed spectrum and the rotation footprint dominates.
This effect gets weaker at a redshift $z=0.1$ and disappears at $z=0.17$ because the spectroscopic fibre probes a larger part of the rotation curve, and a broader emission line profile is observed. 

For gSa + gSa merger simulations, we find an even stronger direction dependency. In fact, we see for all three different redshifts a strong DP fraction at inclinations of $30^{\circ} < \theta < 60^{\circ}$. This is due to the fact that the central disc is more concentrated in comparison to what we see in gSb + gSb simulations. In such a case, a DP signature gets more diluted when viewed edge-on due to gas moving perpendicular to the line-of-sight. At redshift $z=0.05$ and $z=0.1$, it is furthermore very unlikely to observe a DP signature from an edge-on perspective. Only for a redshift of $z=0.17$, we start to detect DP signatures from the edge-on view. 
Since these observations cover a larger surface, gas that is just about to fall back to the central regions is included in the line-of-sight velocity distribution and broadens the emission line.

\subsubsection{The morphology of post-coalescence major mergers}\label{sssect:morph_major}
\begin{figure*}
\centering 
\includegraphics[width=0.98\textwidth]{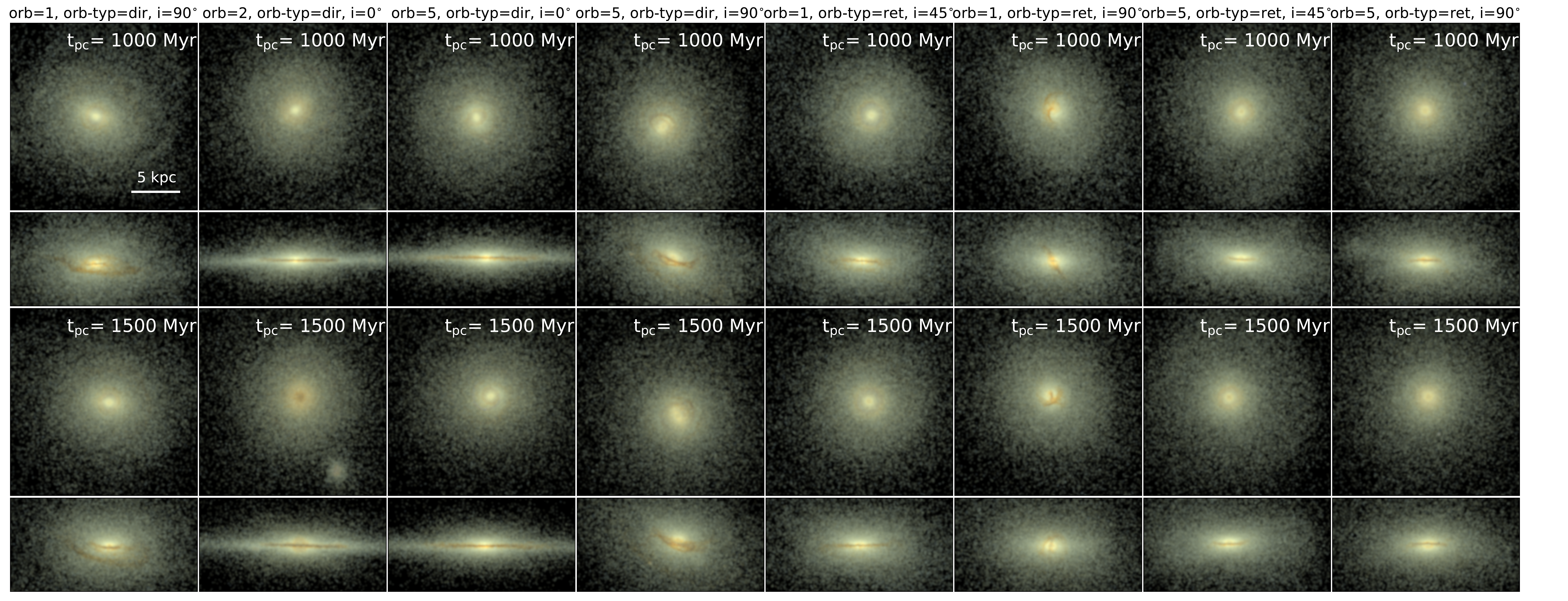}
\caption{Mock $rgb$ snapshots created with $g'$, $r'$ and $i'$ bands of gSb + gSb merger simulations. The images were produced using the radiative transfer software {\sc PEGASE-HR} \citep{2004A&A...425..881L}. We precise the orbital parameters in the title and show below each simulation from the face-on ($40 \times 40\,{kpc}$, $200\times 200 $ pixels) and edge-on ($20 \times 40\,{kpc}$, $100\times 200 $ pixels) perspective at 1000\,Myr and 1500\,Myr after the final coalescence.}
\label{fig:morph_sb}%
\end{figure*}
\begin{figure}
\centering 
\includegraphics[width=0.48\textwidth]{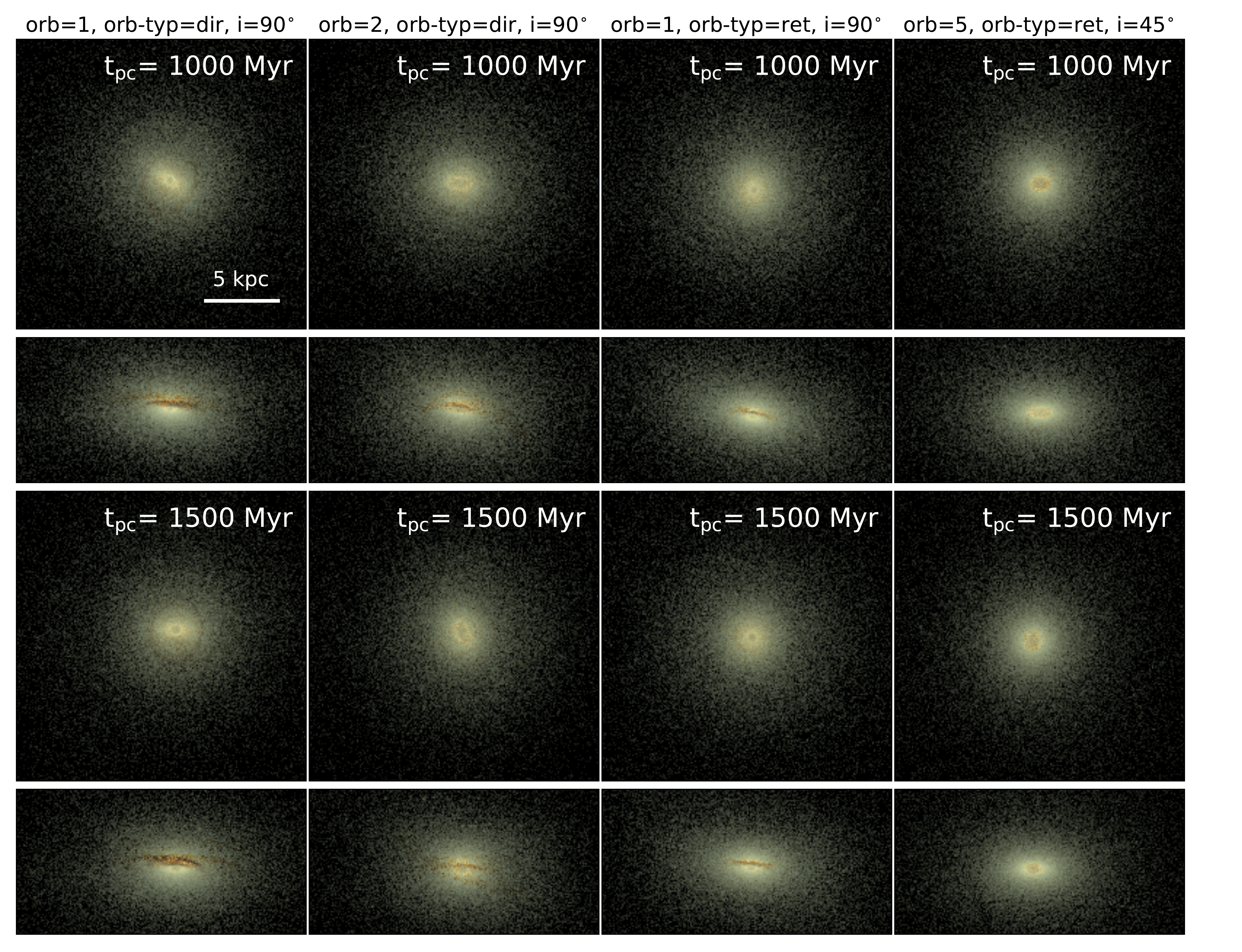}
\caption{Same as Fig.\,\ref{fig:morph_sb}, but for gSa + gSa simulations.}
\label{fig:morph_sa}%
\end{figure}
One of the central results of \citetalias{2020A&A...641A.171M} is that DP emission-line signatures are more likely found in S0 galaxies and in bulge-dominated disc galaxies. Furthermore, no higher merger rate was found in comparison with single-peak emission-line galaxies at the same redshift and with the same stellar mass distribution. 
In order to discuss how relevant major mergers are for the discussion on the origin of DP signatures, two aspects are of particular interest when looking at the morphology: (1) do the post-coalescence mergers still show disc components and (2) can tidal features and merger remnants still be identified with photometric observations?
In order to test this hypothesis, we computed mock $rgb$-images created with the $g'$, $r'$ and $i'$-band filters of the galaxies 1000\,Myr and 1500\,Myr after the final coalescence. 
The broadband colours are computed from stellar population {\sc PEGASE-HR} models \citep{2004A&A...425..881L} which is implemented in the {\sc GalMer} database access\footnote{\url{http://galmer.obspm.fr/}} \citep{2010A&A...518A..61C}. In order to estimate the intensity in each band, light rays are traced along the line-of-sight and attenuation through dust was included. The dust was modelled as explained in \citet{2010A&A...518A..61C}.
In Fig.\,\ref{fig:morph_sb} (resp. \ref{fig:morph_sa}), we show the $rgb$-images for the face-on and edge-on perspective for the gSb + gSb (resp. gSa + gSa) merger simulations. 
In only some snapshots, we are able to identify small tidal features or a miss-aligned dust-lane that can indicate a recent merger. For the majority of snapshots, we observe a smooth morphology. In the two cases of gSb + gSb galaxies with a collision angle of $0^{\circ}$, we even observe a prominent disc. However, as discussed in \citet{2010A&A...518A..61C}, these kinds of orbits are unlikely to happen. For all other simulations, we observe an elliptical galaxy which in some cases still has a disc or has a high ellipticity and is of Hubble type E6 or can be identified as S0, as discussed in \citet{2018A&A...617A.113E}.  

\subsection{Double-peak signatures in minor mergers}\label{ssect:monor_merger}
Since minor mergers are discussed to be responsible for a large fraction of observed DP emission-line galaxies in the literature \citepalias[e.g.][]{2020A&A...641A.171M}, we here discuss how such a kinematic signature can originate from a minor merger event.
We will explore the merger orbits in the same manner as we discussed major mergers (see Sect.\,\ref{ssect:major_merger}) and discuss their morphology.
\subsubsection{Two gas populations detected in one spectra}\label{sssect:dp_minor}
\begin{figure*}
\centering 
\includegraphics[width=0.98\textwidth]{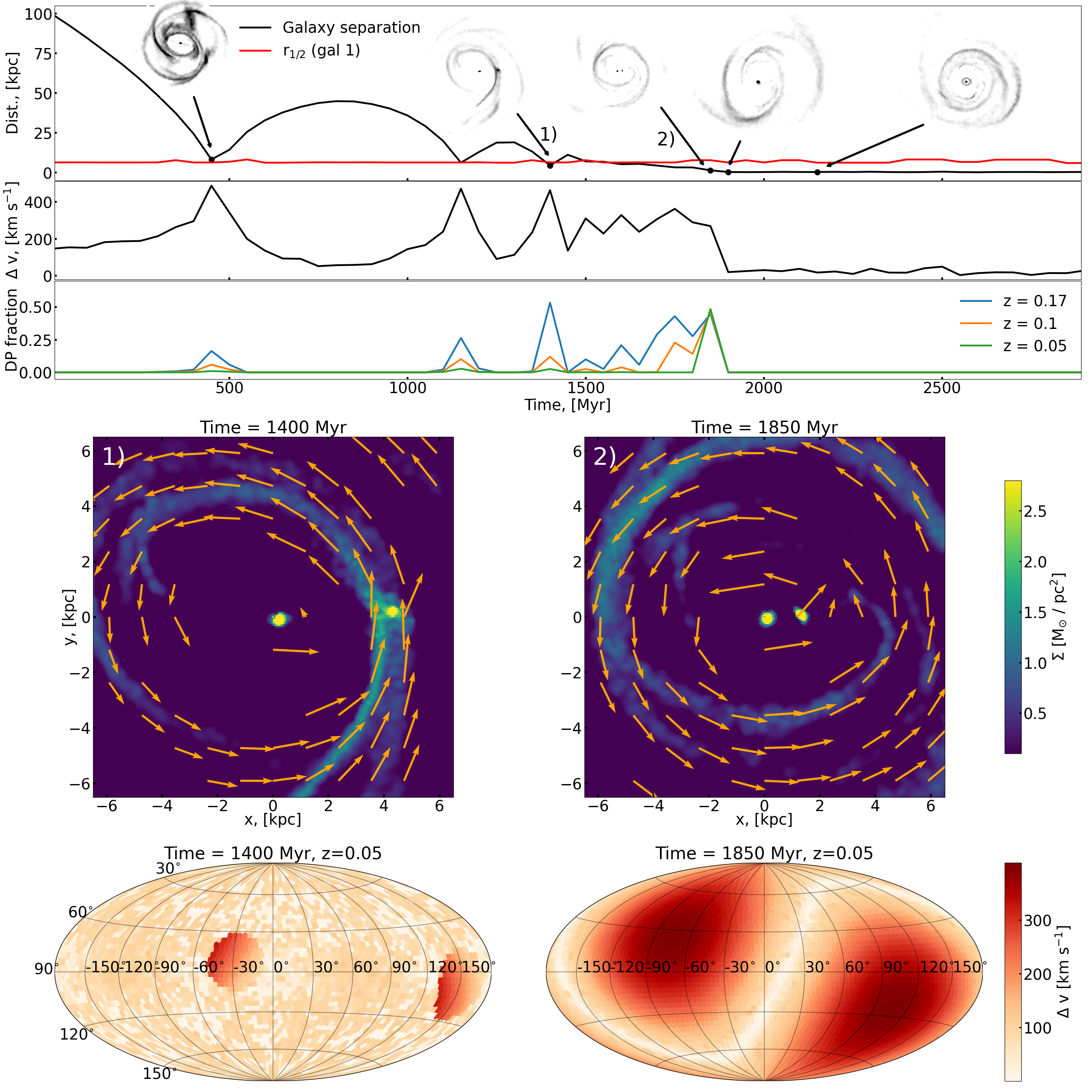}
\caption{Visualisation of a minor merger between a gSa + dSd, simulated with the orbit-id 3 and a direct collision. We show the top panels and two 2D projected gas maps in the same manner as described in Fig.\,\ref{fig:post_coal_sb}. We show two specific snapshots: one close encounter and the moment of final coalescence. The two 2D-projected snapshots are indicated with numbers to better assign them to the orbit, shown on the top. On the two bottom panels, we show direction maps and indicate, with colour maps, the $\Delta v$ measured with the double Gaussian fit to each observed spectrum.}
\label{fig:post_coal_minor}%
\end{figure*}
In order to explore how a minor merger can produce a central DP signature, we compute the directional depending DP fraction of all minor-merger simulations selected in Sect.\,\ref{ssect:merger_parameters}. 
In Fig.\,\ref{fig:post_coal_minor}, we visualise the merger process of a direct (for the giant galaxy) merger encounter between a gSa + dSd and with the orbit-id 3. In comparison to the orbits observed for major-merger simulations (see Sect.\,\ref{ssect:major_merger}), we observe longer merger timescales for minor mergers until the final coalescence. In fact, retrograde (for the giant galaxy) orbits take, for minor-merger simulations, longer than the simulated time span to reach coalescence. Since we are interested in post-coalescence behaviour of galaxies, we selected 6 gSa + dSd and 5 gSa + dSb simulation with direct orbits.

During the merger process, we can clearly identify the two nuclei of the giant and the dwarf galaxies. During close encounters of the two nuclei, we can observe a DP signature. However, only at the final coalescence, where the nuclei of the dwarf galaxy migrates closer than the half mass radius of the giant galaxy, we clearly see a DP signature with observations of the very centre. In this simulation step, the two nuclei are inside the spectroscopic fibre measurements of the $z=0.05$ observation. In Fig.\,\ref{fig:post_coal_minor}, we present the 2D projection of two close encounters and the direction maps indicating from which direction one can observe the highest $\Delta v$ with a double Gaussian fit to the line-of-sight velocity distribution. In the first encounter at 1400\,Myr, the two nuclei are separated by a distance of less than 5\,kpc. We observe a DP signature in more than 50\,\% of the directions at a redshift of $z=0.17$. This is due to the fact that the $3^{\prime\prime}$ spectroscopic fibre covers the central region of 10\,kpc and therefore covers the two nuclei. This is not the case for observations at a redshift of $z=0.05$ and $z=0.1$ and the DP fraction for these observations is significantly smaller. In fact for smaller redshift one can only detect a DP when observing from an angle where both nuclei are covered by the fibre. This is shown in the bottom-left panels of Fig.\,\ref{fig:post_coal_minor}, where we only detect a signal of the dwarf galaxy for a small set of observation angles. On the bottom-right panels of Fig.\,\ref{fig:post_coal_minor}, we show the measured $\Delta v$ for the snapshot where the two nuclei are separated at about 1.5\,kpc before finally merging to one nucleus. A spectroscopic observation at a redshift $z=0.05$ covered both nuclei and a large $\Delta v$ value of up to 400\,km\,s$^{-1}$ can be observed. For this specific case we also observe a DP for observations nearly face-on with an inclination of $\theta \sim 10^{\circ}$. 

\begin{figure}
\centering 
\includegraphics[width=0.48\textwidth]{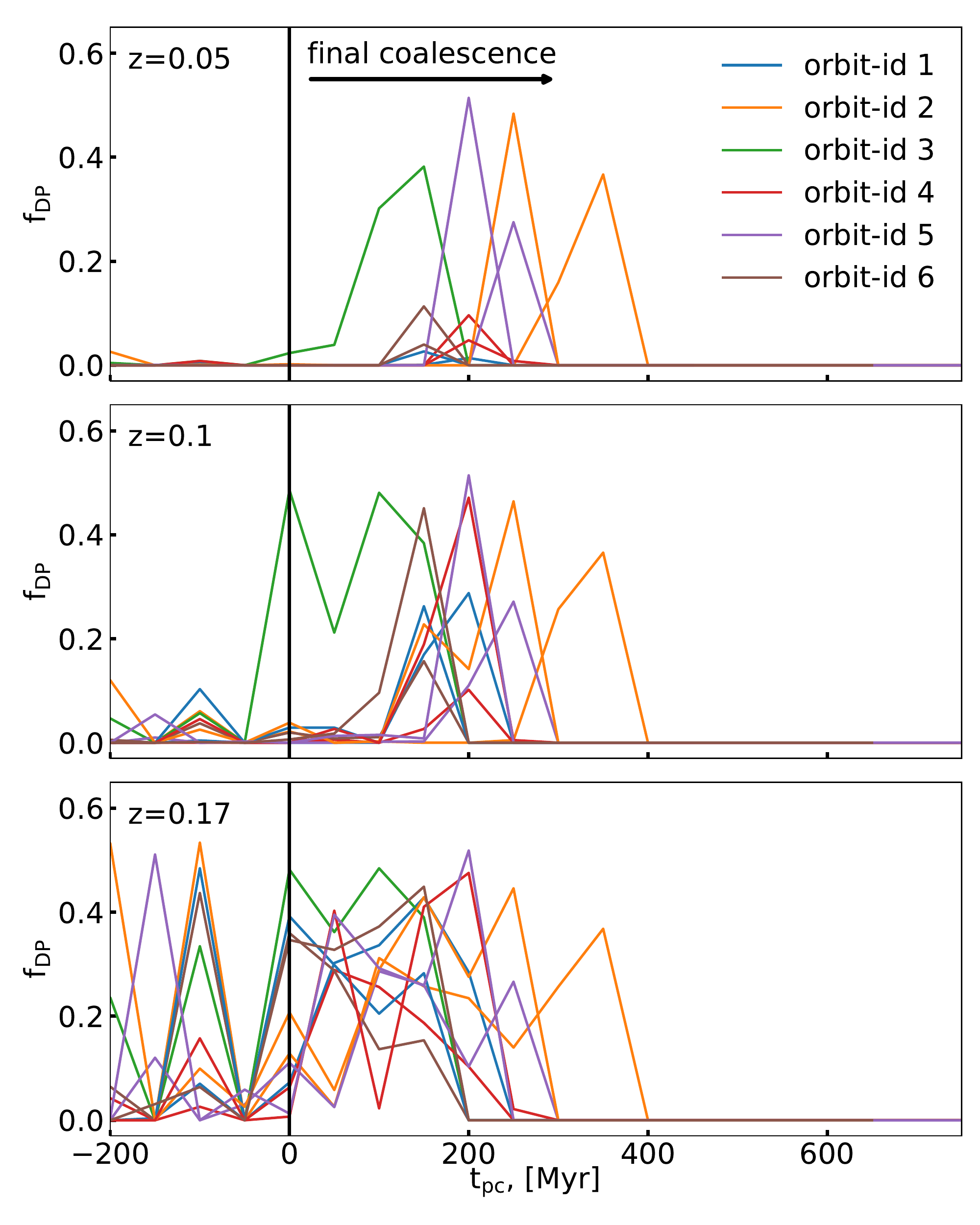}
\caption{Double peak fraction observed at different redshifts after the final coalescence of minor-merger simulation. On the $x$-axis, we show the post-coalescence time ${\rm t_{pc}}$, starting at the moment of final coalescence. The moment of final coalescence is marked by a black line and the values of DP-fraction are indicated on the $y$-axis. The line colour represents the orbit-id specified in Table\,\ref{table:orb_param}.}
\label{fig:orbit_minor}%
\end{figure}
Taking all minor merger observations into account, we can see a clear pattern: at close encounters, we find higher DP fractions. However, for the redshift of $z=0.05$, the value is largest in the closest configuration directly before the two nuclei finally merge. After the final coalescence, no DP can be detected and no rotating disc as seen in major mergers is formed. In Fig.\,\ref{fig:orbit_minor}, we show the DP fraction of all minor-merger simulations at the final coalescence observed at different redshifts. As described in Sect.\,\ref{ssect:merger_characteristics}, the final coalescence is defined by the moment where the COMs of the two galaxies approach each other less than the half mass radius of the giant galaxy without subsequently moving away from each other. For the redshift $z=0.17$, we observe DP signatures between 50 to 350\,Myr after the final coalescence. This is in all cases the moment when the two nuclei are closer than the spectroscopic fibre size. Since for the redshift at $z=0.1$ and $z=0.17$ the fibre size is larger we see in these observations a DP signature earlier in the final coalescence. However, the moment of the last detected signature does not depend on the redshift. This is due to the fact that the two nuclei finally merge. By observing the directions of detection of DP at the final detection, we do not see a preferred observation direction and a DP can be seen from a face-on view in some cases.

\subsubsection{The morphology of minor mergers at the final coalescence}\label{sssect:morph_minor}
\begin{figure*}
\centering 
\includegraphics[width=0.98\textwidth]{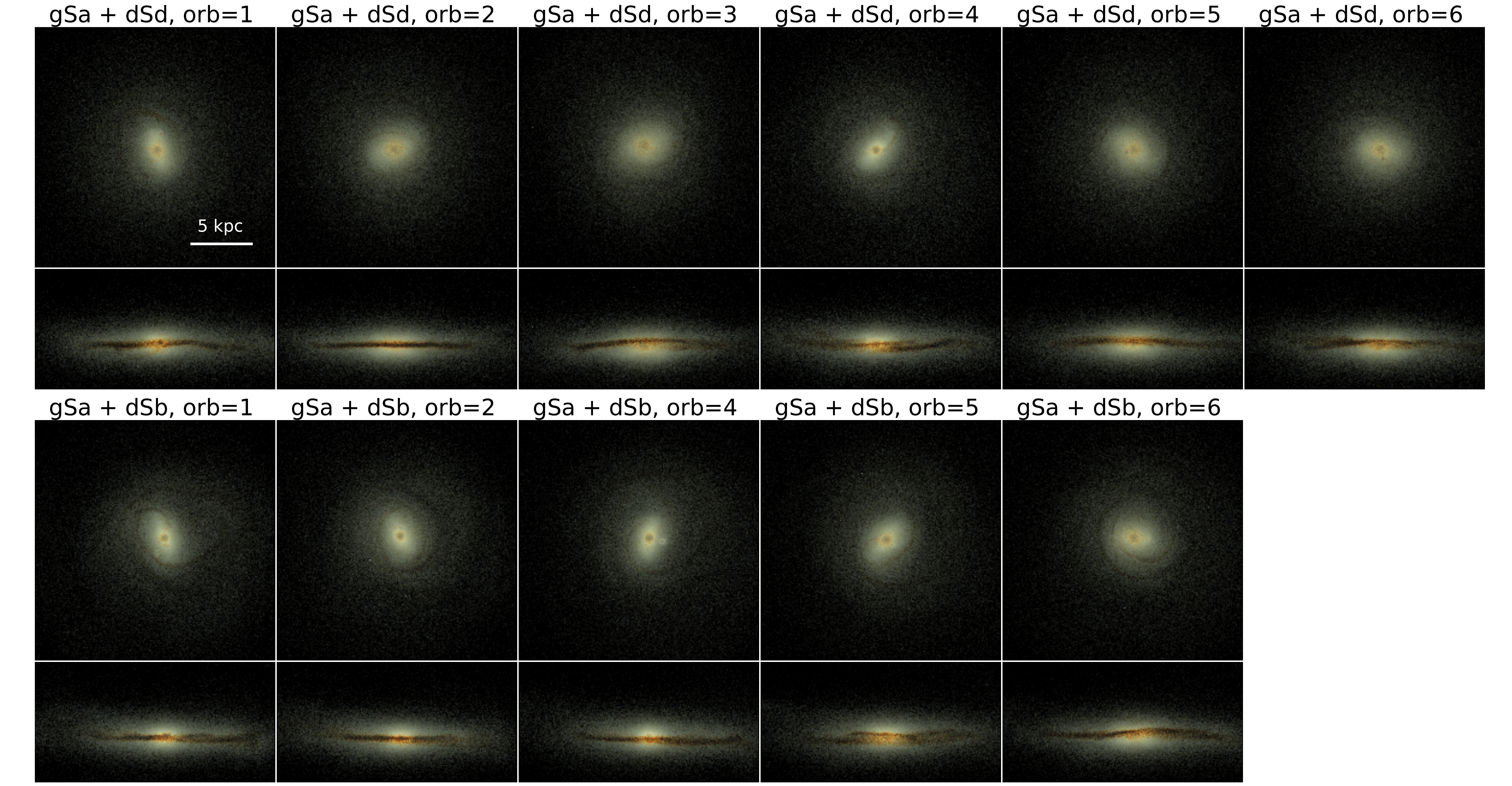}
\caption{Mock $rgb$-images of gSa + dSd and gSa + dSb merger simulations computed in the same manner as in Fig.\,\ref{fig:morph_sb}}
\label{fig:morph_minor}%
\end{figure*}
At the moment of final coalescence in major mergers, one can find high DP fractions similar to minor mergers. However, there is a big difference: major mergers show very strong perturbations at this moment which is easy to identify even at a higher redshift. 
Minor mergers, on the other hand, are not known to have such a strong impact on the morphology. In Fig.\,\ref{fig:morph_minor}, we present the morphology of all the minor-merger simulations after the final coalescence when the largest DP fraction at $z=0.05$ is measured. In only one case, the two nuclei can be clearly identified from the face-on, although this would only be visible with high-resolution images. In all post-coalescence minor mergers, we see a nearly undisturbed disc structure and it would be difficult to distinguish such a galaxy from an isolated galaxy. 

\section{Discussion}\label{sect:discussion}
\subsection{Double-peak signatures from rotating discs: isolated galaxy vs. post merger}\label{ssect:disc_rot_vs_merger}
A spectroscopic observation of an entire rotating gaseous disc is known to describe a double-horn profile, when observed inclined \citep[e.g.][]{2014MNRAS.438.1176W}. However, this is well known for the HI line, measured for an entire galaxy. Ionised gas kinematics in the centre of a galaxy, on the other hand, traces only an inner small part of the rotation curve. Massive bulges in disc galaxies are known to create a strong velocity gradient in the central region \citep{2001ARA&A..39..137S}. 
Using axisymmetric models of discs with pure rotation, we find that the DP signature primarily depends on the angle of observation: the higher the inclination, the larger the separation of a double-Gaussian fit. Furthermore, we find the strongest DP signatures for high bulge concentrations when only observing the central 3\,kpc.
This analytical view points out one aspect quite clearly: DP emission lines have a strong connection to the bulges of galaxies. Accordingly, massive or highly concentrated bulges in galaxies can create a sufficient deep gravitational potential to cause high velocity gradients at the centre.

In Sect.\,\ref{ssect:major_merger}, we find that  a centralised disc can be formed in a late stage of a post major galaxy merger. 
Major mergers generally destroy the disc morphology of the two progenitors and result in an elliptical galaxy as demonstrated in \citet{1982ApJ...259..103F, 1983MNRAS.205.1009N}. These findings were further confirmed for dry major mergers \citep{2020MNRAS.493.1375P}. For gas rich major mergers however, a disc can be formed in the post merger phase from a gaseous disc that subsequently re-settles \citep{2009MNRAS.398..312G, 2009ApJ...691.1168H}.
In violent major mergers which undergo a phase of ultra-luminous infra-red emission, a centralised molecular gas disc was detected in \citet{1998ApJ...507..615D}. \citet{2009A&A...493..899P} reported a gas rich disc which might be the result of a collapse of a larger disc or a major merger. 

After the final coalescence of a major merger, in-falling gas from tidal tails can form a rotating gaseous disc over long periods of time \citep[][]{2002MNRAS.333..481B}. In the major-merger simulations which we consider in this work, we do see this behaviour: at about 1000\,Myr after the final coalescence, gas which was slung far outside the merging system due to tidal tails formed a central disc, which we observe as a double-peak emission line. We do see stronger DP signature for the resulting galaxy of a gSa + gSa galaxy merger in comparison to a gSb + gSb. This is most likely due to the progenitors of the latter having less massive bulges and the resulting rotation curve shows smaller velocities in the centre.

Regarding the morphology of the late stages of major mergers, we observe indeed mostly early type morphologies. Only in mergers with the two discs in the orbital plane do we observe a prominent disc structure in the resulting galaxy. However, besides such mergers, we do not find any dependencies on the orbital geometry of the merger, which was further discussed in \citet[][]{1996ApJ...464..641M}.
At the observed stage of post-coalescence, we do not observe strong tidal features in the central kiloparsecs, which is in line with the findings from \citet{2010MNRAS.404..575L}. 
In this work, we only consider visual merger identification similar to e.g. \citet{2018MNRAS.476.3661D} or \citet{2013MNRAS.435.2835W}. This is only sensitive to prominent tidal features and perturbations which are detectable in early merger stages. The detection of post-coalescence galaxy mergers are difficult to detect and are often accompanied by large bulges \citep[e.g.][]{1991ApJ...370L..65B, 1992ApJ...393..484B} or dual nuclei \citep[e.g.][]{2003ApJ...582L..15K}. However, the presence of large bulges do not give any insight concerning the merger time-scale, and in order to identify dual nuclei, high-resolution observations are needed.
By combining multiple imaging predictors to a linear discriminant analysis method, it is possible to correctly identify post-coalescence galaxy mergers as shown with simulated galaxies by \citet{2019ApJ...872...76N}. When including stellar kinematics, observed with integrated-field spectroscopy, the post-coalescence mergers can be even better identified \citep{2021ApJ...912...45N}. 
In a similar work on galaxy mergers from cosmological simulations, \citet{2022MNRAS.511..100B} have shown that accurate identification can also be achieved with neuronal networks, even though they find that the kinematic input has a less significant contribution compared to the imaging input.

For the post coalescence major mergers in this work, we find morphologies indicate a strong ellipticity so that the galaxies can be identified as lenticular galaxies. This was discussed for the same merger simulations in \citet{2018A&A...617A.113E}. 
These galaxies can correspond to the excess of DP S0 galaxies, found in \citetalias{2020A&A...641A.171M}. However, these configurations of post-major mergers form around 1\,Gyr after the final coalescence. 
The increased star-formation rate associated with a merger has already faded away at this merger stage \citep{1996ApJ...464..641M, 2007A&A...468...61D}. Hence, this is in conflict with the increased star-formation rates found for DP galaxies \citep{2019A&A...627L...3M, 2020A&A...641A.171M, 2021arXiv211212796M}.

\subsection{Strong double-peak features in disc galaxies: bars or minor mergers ?}\label{ssect:bar_vs_minor}
In contrast to major mergers, minor mergers are less violent and the merger morphology is detectable only up to 100\,Myr after the final coalescence using photometry \citep[e.g.][]{2010MNRAS.404..575L}. In addition to that, within this timescale an enhance star-formation rate can be induced by the merger \citep{2007A&A...468...61D}. 
Considering the excess of S0 and Sa galaxies and the central star formation enhancement found in star-forming DP galaxies \citepalias{2020A&A...641A.171M}, a minor merger can explain the observed characteristics of DP emission-line galaxies. 
During a close encounter in a minor merger, one can observe a DP signature which is similar to the case discussed in \citet[][]{2021A&A...653A..47M}. However, this is not necessarily the final stage of the merger but a superposition of two galaxies aligned with the line-of-sight. This phenomenon will be addressed in greater detail in Halle et al. (in prep.). In this work, however, we set the focus on how to create a DP signature which cannot be identified though visual inspection. 
Depending on the merger orbit, the dwarf galaxy can enter from any direction into the central region and we therefore do not detect any directional dependence. In some cases, we even observe a strong DP signature from a face-on perspective. 
Within 350\,Myr, the two nuclei finally merge and no DP signature can be detected anymore. In the merger stage of final coalescence, when we observe the highest DP fraction, we only see weak tidal features in the central kiloparsecs. The two nuclei would be only visible with high resolution imaging or in very nearby galaxies. furthermore, minor mergers are considered to happen more frequently than major mergers in the late universe \citep{2005ApJ...620..564C, 2007ApJ...660L..43N}.

As mentioned in Sect.\,\ref{ssect:disc_rot_vs_merger}, a bar feature in isolated galaxies can also create a DP emission line. 
In Sect.\,\ref{ssect:bars}, we explore bars in simulated Sa and Sb galaxies and find a DP when viewing from a perspective parallel to the bar.
Barred galaxies are considered to be effective in transporting cold gas inwards, leading to central growth and rejuvenation of SF in the central region \citep{2019MNRAS.484.5192C}. On the one hand, minor mergers can trigger a central star-formation enhancement \citep[e.g.][]{2014MNRAS.438.1870D}, on the other hand, however, bars are considered to trigger central star bursts more effectively than galaxy-galaxy interactions \citep{2011MNRAS.416.2182E}. 
Therefore, also barred galaxies would be a considerable mechanism to produce a strong DP emission-line signature accompanied by a central star-formation enhancement.

Observing a bar parallel to its major axis can lead to a false classification of a disc galaxy with a symmetric bulge. In fact, \citetalias{2020A&A...641A.171M} finds only 3\,\% barred galaxy for DP emission-line galaxies. However, the used identification of this galaxy type is favouring less inclined and face-on galaxies as they are detected with a machine-learning algorithm described in \citet[][]{2018MNRAS.476.3661D}. Thus, a large part of more inclined barred galaxies might be not detected, the bar being hidden due to the viewing angle. 
In principle, bars can occur in spiral and S0 galaxies. These types make up half of the \citetalias{2020A&A...641A.171M}-DP galaxy sample (16\,\% spiral and 36\,\% S0 galaxies). By combining a bar fraction from observations and the estimated DP fraction due to a bar, we can estimate whether bars can be responsible alone for the significant increase in DP S0 galaxies observed by \citetalias{2020A&A...641A.171M}. 
We adopt a bar fraction of spiral galaxies of ${\rm P(bar| spiral)} = 0.66$ \citep{2000AJ....119..536E} and for S0 galaxies of ${\rm P(bar| S0)} = 0.46$ \citep{2009ApJ...692L..34L}. 
In order to estimate the frequency of S0 and spiral galaxies, we use the same morphological selection of these galaxy types as performed in \citetalias{2020A&A...641A.171M}, based on \citet{2018MNRAS.476.3661D}, for SDSS galaxies. We further restrain the selection to a similar stellar mass and redshift distribution by applying a stellar-mass cut of ${\rm M_*} \geq 10^{10.5}\, {\rm M_{\odot}}$ and a redshift cut of $z \leq 0.2$. 
This selection results in fractions of P(spiral) = 0.131 and P(S0) = 0.255.  
The fact that we find more S0 galaxies is due to the selection of galaxies with high stellar masses which is similar to the selection in \citetalias{2020A&A...641A.171M}. Since we assume a lower probability of S0 galaxies exhibiting a bar (${\rm P(bar| S0)}$), it is rather unlikely that the factor $\sim 2$ we see in the ratio between S0 and spiral galaxies in the \citetalias{2020A&A...641A.171M}-DP sample can be explained purely by bars. However, this estimation is based on two simplified assumptions: first, that bars in S0 galaxies produce the same DP fraction as in spiral galaxies and second, that the bar fraction is constant for all stellar masses $\geq 10^{10.5}\, {\rm M_{\odot}}$, which is not the case \citep{2020ApJ...895...92Z,2020ApJ...904..170Z,2021MNRAS.508..926R}. 
In this paper we address the fundamental question of which mechanisms can cause DP signatures. However, it is difficult to estimate which of these effects is more likely based on idealised simulations and we therefore plan to estimate this question in a future work.

\subsection{Resolving double-peak emission lines and the importance of future surveys}
Here, we discussed multiple mechanisms which can lead to a DP signature observed in a central spectroscopic observation of a galaxy. However, considering only a central spectrum and a snapshot in the optical light, one cannot conclusively determine the origin of the DP emission line. In order to distinguish between the different mechanisms, discussed here, additional information about a spatial distribution of the kinematic signatures is needed. 

As shown in \citet[][]{2021A&A...653A..47M}, relying on integrated field spectroscopy with the Mapping Nearby Galaxies at APO \citep[MaNGA,][]{2015ApJ...798....7B} survey, one can spatially disentangle two different gas components. In this very case, the central DP signature found in the central $3^{\prime\prime}$ SDSS spectrum, originates from two superposed discs. Long-slit spectroscopic observations provide a spatial resolution and \citet[e.g.][]{2009ApJ...702L..82C, 2011ApJ...737L..19C, 2015ApJ...813..103M, 2016ApJ...832...67N} and \citet{2018ApJ...867...66C} succeeded in resolving a dual AGN as the underlying mechanism of a detected DP and distinguished them from other mechanisms such as gas outflows or rotating discs. 
Therefore, the mechanisms, discussed in this work should be studied in greater detail by means of surveys such as MaNGA, but at the same time the basic understanding of these phenomena should be investigated with further simulations. Cosmological simulations, in particular, offer a special opportunity as they provide a much greater diversity of different merger scenarii, and galaxies are in constant interaction with their environment. Furthermore, the inclusion of AGN feedback allows to even further discuss how DP emission lines are connected to physical processes \citep[][]{2015ARA&A..53...51S, 2020NatRP...2...42V}. A complete analysis of SDSS-like spectroscopic observations in cosmological simulations may also provide insight into which underlying process is more likely (e.g. bar signatures or minor merger).

By aiming for a better understanding of the kinematic footprint of gas in galaxies, we might also be able to apply such insights to upcoming surveys. The SDSS only observed galaxies in the late Universe. By using DP emission-line signatures as a tracer to study gaseous discs and galaxy mergers, we can better estimate the merger rate over larger ranges of redshift. This would help us to understand for example how galaxies evolve through mergers and quantify how the star-formation rate is connected to such phenomena. 
Two upcoming surveys are of special interest for this very task: the VLT 4MOST survey as it probes emission-line galaxies up to a redshift of $z=1.1$ \citep[][]{2019Msngr.175...50R} and the EUCLID mission which will provide spectroscopic data for galaxies up to $z\sim2$ \citep[][]{2011arXiv1110.3193L}. Spectroscopic observations from the EUCLID mission will not be able to resolve DP signatures due to the insufficient spectral resolution of R=250 at a pixel size of $0.3^{\prime\prime}$, however, the high imaging resolution of $0.1^{\prime\prime}$ will enable to probe earlier stages of merger with an unprecedented sample size.
Visual galaxy mergers and DP emission-line galaxies can be used as a tool to select promising candidates in the high redshift universe and compare the measured kinematic footprint and merger rate to the ones we know from the late Universe.

\section{Conclusions}\label{sect:conclusion}
A double-peak (DP) emission line, observed in the centre of a galaxy is a peculiar feature, as it offers insights into the central kinematic processes. This kinematic footprint has been used to find dual active galactic nuclei (AGN) or AGN-driven gas outflows. In recent studies, a broader search for DP galaxies has been conducted in order to shed light on this phenomenon from a more general perspective. The resulting DP sample showed that AGNs represent only a small subgroup and the majority shows only moderate or no AGN activity. 
Furthermore, DP galaxies are predominantly S0 or disc galaxies with large bulges and no increased merger rate was observed. Taking into account that star-forming DP galaxies exhibit a central star-formation enhancement, the most plausible explanation would be the observation of a minor merger. However, without followup observations one cannot conclusively determine the underlying mechanism for an individual galaxy. 

In order to get a better understanding of the internal kinematic processes creating a DP signatures, we investigated different possibilities in this work. 
We, therefore, computed synthetic SDSS spectroscopic emission-line observations from disc models and simulations and searched for DP signatures from all directions using a grid of observation angles.
With axisymmetric models, we explored from which observation angle and for which rotation curves one can see a DP. To get a more realistic view, we searched in simulations of isolated galaxies from where we can observe a DP signature and found besides a rotation pattern that bars can create a strong DP when observed parallel to the major axis of the bar.
We also observed minor and major-merger simulations over the course of their merger process. We found DP signatures during close encounters of two galaxies as two gas components are present inside the spectroscopic observation.
Furthermore, after about 1\,Gyr after the final coalescence, we see a central rotating disc in post-major mergers which create a distinct DP fraction.
This phenomena however, is not detected in minor-merger simulations. However, a strong DP signature is observed within 350\,Myr after the final coalescence. 
For the discussed stages of major and minor merger simulations, the morphology does not give a direct indication of a recent merger.

Using axisymmetric models, we have gained a clear understanding of how the connection between the stellar bulge and the rotation curve can lead to a DP. Massive or highly concentrated bulges can create a strong central velocity gradient such that a DP can be observed at low inclinations of $\theta = 40^{\circ}$ ($\theta = 0^{\circ}$ would be face-on). 
In the context of observed DP galaxies in the SDSS, we must clearly say that late cycles of major mergers are unlikely, as they tend to produce S0 and mainly elliptical morphologies. Moreover, at the merger stage, we discuss here, they have already consumed the majority of their gas for star formation and an enhanced star-formation rate is close to impossible. 
Minor mergers and bars as a mechanism for DP signatures show great agreement with observations. On the one hand, both are known for central star-formation activity and, on the other hand, both phenomena occur frequently. Although the range in which we can observe a DP in minor mergers is relatively short (about 350\,Myr), however, this footprint can be seen from a large range of angles and there is no correlation with the galaxy inclination. 
These findings show further possibilities of how one can interpret an observed DP emission line. And at the same time it is in line with the observations of which minor mergers were discussed as the most plausible explanation.

In the context of future work on DP emission-line galaxies, we further discussed that using integrated-field spectroscopy can disentangle the underlying mechanisms. Furthermore, the understanding of DP emission lines is a crucial tool for upcoming spectroscopic surveys at high redshift, as they can help to identify galaxy mergers. 

\begin{acknowledgements}
This work was supported by the {\em Programme National Cosmology et Galaxies} (PNCG) of CNRS/INSU with INP and IN2P3, co-funded by CEA and CNES. IC's research is supported by the SAO Telescope Data Center. IC acknowledges support from the RScF grant 19-12-00281.
\end{acknowledgements}

\bibliographystyle{aa}
\bibliography{Mybiblio}

\appendix

\section{Initial galaxy parameters}\label{app:init_param}
In this section we provide detailed parameters of galaxy simulations of the {\sc GalMer} project, described in Sect.\,\ref{sssect:sim_design} and \ref{ssect:merger_parameters}. Table\,\ref{table:init_params} summarises the initial parameters of all individual galaxy types used in this work and Table\,\ref{table:orb_param} summarises orbital parameters of the merger simulations.
\begin{table}[h!]
\caption{Initial parameters of simulated galaxies in the {\sc GalMer} database.}
\label{table:init_params}
\begin{tabular}{l c c c c }
\multicolumn{1}{l}{} & gSa & gSb & dSb & dSd \\
\hline
$M_{\rm gas} [2.3\times 10^9 M_{\odot}]$  & 4 & 4 & 0.4 & 0.75 \\
$M_{\rm *\, disc} [2.3\times 10^9 M_{\odot}]$  & 40 & 20  & 2 & 2.5 \\
$M_{\rm *\, bulge} [2.3\times 10^9 M_{\odot}]$  & 10 & 5 & 0.5 & 0 \\
$M_{DM} [2.3\times 10^9 M_{\odot}]$  & 50 & 75 & 7.5 & 7.5 \\
$a_{\rm gas}$ [kpc] & 5 & 6 & 1.6 & 2.2 \\
$h_{\rm gas}$ [kpc] & 0.2 & 0.2 & 0.06 & 0.06 \\
$a_{\rm *, disc}$ [kpc] & 4 & 5 & 1.6 & 1.9 \\
$h_{\rm *, disc}$ [kpc] & 0.5 & 0.5 & 0.16 & 0.16 \\
$b_{\rm *, bulge}$ [kpc] & 2 & 1 & 0.3 & - \\
$b_{\rm DM}$ [kpc] & 10 & 12 & 3.8 & 4.7 \\
\end{tabular}
\tablefoot{The values are taken from \citet{2010A&A...518A..61C}.}
\end{table}
\begin{table}[h!]
\caption{Orbital parameters for major and minor mergers used in the {\sc GalMer} database.}
\label{table:orb_param}
\begin{tabular}{l c c c c}
orb.id & r$_{\rm ini}$ & v$_{\rm ini}$ & L &  spin \\
 & kpc & 10$^2$\,km\,s$^{-1}$ & 10$^2$\,km\,s$^{-1}$\,kpc & \\
\hline
\multicolumn{5}{c}{Major merger}\\
\hline
01dir & 100 & 2. & 56.6 & up  \\
01ret & 100 & 2. & 56.6 & down \\
02dir & 100 & 3. & 59.3 & up  \\
02ret & 100 & 3. & 59.3 & down \\
03dir & 100 & 3.7 & 62.0 & up  \\
03ret & 100 & 3.7 & 62.0 & down \\
04dir & 100 & 5.8 & 71.5 & up  \\
04ret & 100 & 5.8 & 71.5 & down \\
05dir & 100 & 2. & 80.0 & up  \\
05ret & 100 & 2. & 80.0 & down \\
\hline
\multicolumn{5}{c}{Minor merger}\\
\hline
01dir & 100 & 1.48 & 29.66 & up  \\
01ret & 100 & 1.48 & 29.66 & down \\
02dir & 100 & 1.52 & 29.69 & up  \\
02ret & 100 & 1.52 & 29.69 & down \\
03dir & 100 & 1.55 & 29.72 & up  \\
03ret & 100 & 1.55 & 29.72 & down \\
04dir & 100 & 1.48 & 36.33 & up  \\
04ret & 100 & 1.48 & 36.33 & down \\
05dir & 100 & 1.52 & 36.38 & up  \\
05ret & 100 & 1.52 & 36.38 & down \\
\end{tabular}
\tablefoot{The values are taken from \citet{2010A&A...518A..61C}.}
\end{table}

\section{Merger orbit of major merger galaxies}\label{sect:spectra}
\begin{figure*}
\centering 
\includegraphics[width=0.98\textwidth]{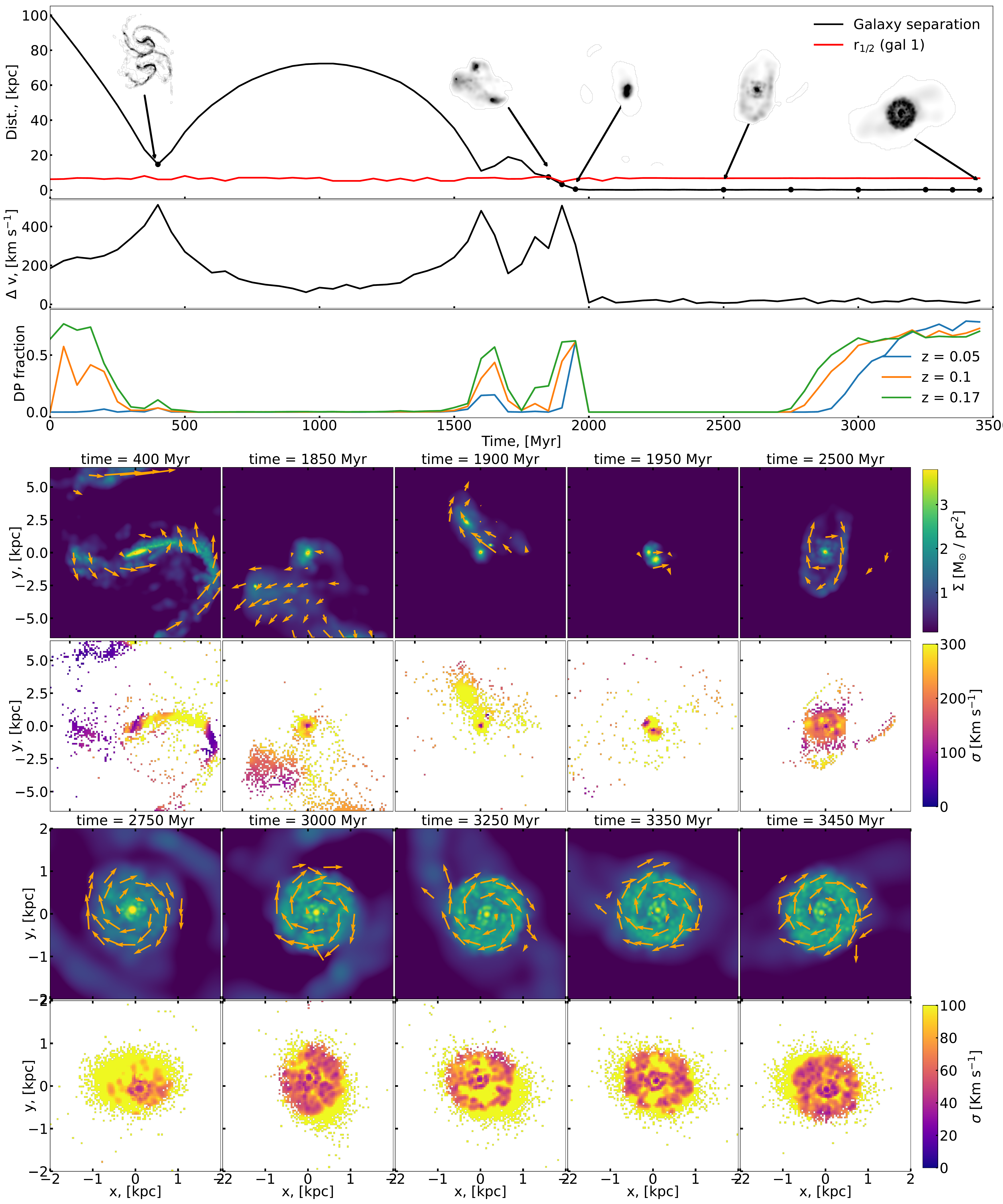}
\caption{Visualisation of a gSa + gSa galaxy merger. The presentation is the same as described in Fig.\,\ref{fig:post_coal_sb}. However, in the bottom panels, presenting the 2D projection of different snapshots, we display the central 4\,kpc to better visualise the central disc. The merger process is characterised by an retrograde orbit with the orbit-id 5 and an inclination of $45^{\circ}$.}
\label{fig:post_coal_sa}%
\end{figure*}

In this section, an additional figure of a major merger simulation of two gSa galaxies is presented in Fig.\,\ref{fig:post_coal_sa}. This is supplementary to Fig.\,\ref{fig:post_coal_sb} which is used to discuss a major merger simulation.

\end{document}